\documentclass[aps,prb,twocolumn,superscriptaddress]{revtex4-2}
\usepackage{natbib}
\usepackage{graphicx}
\usepackage{latexsym}
\usepackage{amssymb}
\usepackage{amsmath}
\usepackage{amsfonts}
\usepackage{dsfont}
\usepackage{gensymb}
\usepackage{bm}
\usepackage{color}
\usepackage{hyperref}
\hypersetup{
colorlinks = true,
linkcolor = [rgb]{0.70,0.13,0.13},
citecolor = [rgb]{0.13,0.55,0.13},
urlcolor = [rgb]{0.25, 0.41, 0.88}}
\usepackage{braket}
\usepackage{physics}
\usepackage{bbold}
\usepackage[caption=false]{subfig}
\usepackage[dvipsnames]{xcolor}
\usepackage[normalem]{ulem}
\usepackage[force]{feynmp-auto}
\usepackage{mathtools}
\usepackage{lipsum}
\usepackage{upgreek}
\usepackage{multirow}
\usepackage{enumitem}
\usepackage{slashed}
\usepackage{makecell}

\newcommand{\ii}{\mathsf{i}}

\newcommand{\phov}[1]{\fmfdot{#1}}
\newcommand{\bosv}[1]{\fmfv{d.sh=square,d.f=shaded,d.si=4}{#1}}
\newcommand{\intv}[1]{\fmfv{d.sh=circle,d.f=empty,d.si=4}{#1}}
\newcommand{\bsv}[1]{\fmfv{d.sh=cross,d.si=6}{#1}}

\newcommand{\la}{\lambda}
\newcommand{\La}{\Lambda}
\newcommand{\ka}{\kappa}
\newcommand{\si}{\sigma}

\newcommand{\ep}{\epsilon}
\newcommand{\Ga}{\Gamma}
\newcommand{\ga}{\gamma}
\newcommand{\be}{\beta}
\newcommand{\al}{\alpha}
\newcommand{\reg}{\ln(k/\La) +reg.}
\newcommand{\bpsi}{\bar{\psi}}

\newcommand{\U}{\mathsf{U}}
\newcommand{\SU}{\mathsf{SU}}
\newcommand{\SO}{\mathsf{SO}}
\newcommand{\gU}{\mathsf{U}}
\newcommand{\gSU}{\mathsf{SU}}

\newcommand{\Oo}{\mathsf{O}}
\newcommand{\Z}{\mathbb{Z}}
\newcommand{\cD}{\mathcal{D}}

\newcommand{\cM}{\mathcal{M}}

\newcommand{\mM}{\mathbf{M}}
\newcommand{\mF}{\mathbf{F}}
\newcommand{\tmF}{\mathbf{\tilde{F}}}
\newcommand{\CX}{\mathcal{X}}
\newcommand{\CL}{\mathcal{L}}
\newcommand{\IZ}{\mathbb{Z}}
\newcommand{\sg}[1]{\sigma^{#1}}
\newcommand{\TT}{{\intercal}}
\newcommand{\tpsi}{\tilde{\psi}}
\newcommand{\cfr}[1]{e^{\ii\frac{\pi}{4}\sigma^{#1}}}

\newcommand{\fpic}[1]{\frac{#1}{4\pi}}
\newcommand{\tpi}{\frac{1}{2\pi}}
\newcommand{\fpi}{\frac{1}{4\pi}}
\newcommand{\ta}{{\tilde{a}}}
\newcommand{\kaequal}{\stackrel{\kappa \rightarrow 0}{=}}

\newcommand{\photonprop}[2]{\Pi_{#1 #2}(q)}
\newcommand{\photonvert}[1]{\ii \mathds{1}_{N_f}\otimes \gamma^{#1}}
\newcommand{\bosonprop}[2]{D(q)\delta_{#1 #2}}
\newcommand{\bosonvert}[1]{M^#1\otimes \mathds{1}_2}
\newcommand{\fermionprop}[2]{-\ii\frac{(#1)_#2(\mathds{1}_{N_f}\otimes \gamma^#2)}{(#1)^2}}

\newcommand{\idm}{m}
\newcommand{\trima}{m'}
\newcommand{\idty}[1]{#1}

\newcommand{\eq}[1]{\begin{equation}#1\end{equation}}
\newcommand{\eqs}[1]{\begin{equation}\begin{split}#1\end{split}\end{equation}}
\newcommand{\eqnref}[1]{Eq.\,\eqref{#1}}
\newcommand{\figref}[1]{Fig.\,\ref{#1}}

\newcommand{\secref}[1]{Sec.\,\ref{#1}}
\newcommand{\appref}[1]{App.\,\ref{#1}}
\newcommand{\refcite}[1]{Ref.\,\onlinecite{#1}}

\newcommand{\vv}[1]{\langle #1 \rangle}
\newcommand{\vect}[1]{{\bm{#1}}}
\newcommand{\dsi}{\mathds{1}}
\newcommand{\ntqed}{$N_f=2$ QED$_3$ }

\newcommand*{\citen}[1]{%
  \begingroup
    \romannumeral-`\x 
    \setcitestyle{numbers}%
    \cite{#1}%
  \endgroup
}

\hypersetup{
  pdfauthor = {Da-Chuan Lu},
}

\begin{document}
\title{Self-duality protected multi-criticality in deconfined quantum phase transitions}
\author{Da-Chuan Lu}
\affiliation{Department of Physics, University of California, San Diego, California 92093, USA\looseness=-1}
\author{Cenke Xu}
\affiliation{Department of Physics, University of California, Santa Barbara, California 93106, USA\looseness=-1}
\author{Yi-Zhuang You}
\affiliation{Department of Physics, University of California, San Diego, California 92093, USA\looseness=-1}
\date{\today}
\begin{abstract}
Duality places an important constraint on the renormalization group flows and the phase diagrams. For self-dual theories, the self-duality can be promoted as a symmetry, this leads to the multi-criticalities. This work investigates a description of the deconfined quantum criticality, the $N_f=2$ QED$_3$, as an example of self-dual theories and its multi-critical behavior under perturbative deformations. The multi-criticality is described by the theory with Gross-Neveu couplings and falls in a different universality class than the standard deconfined quantum criticality. We systematically calculate the scaling dimensions of various operators in the 3d quantum electrodynamics with the Chern-Simons term and Gross-Neveu couplings by the large-$N$ renormalization group analysis. Specifically, we find certain non-relativistic four-fermion interactions which correspond to the dimer-dimer interactions in the lattice model will drive the deconfined quantum criticality to the first-order transition, this result is consistent with previous numerical studies.

\end{abstract}
\maketitle

\section{Introduction}

Duality plays an important role in relating different phases of matter. One famous example is the Kramers-Wannier duality\cite{kramers_statistics_1941} in (1+1)D transverse field Ising model $H=\sum_{i} -J Z_{i}Z_{i+1}-hX_{i}$, which exchanges $J$ and $h$ and maps the ferromagnetic (Ising symmetry breaking) phase to the paramagnetic (Ising symmetric) phase and vice versa. More generally, two theories are dual to each other when they have different ultraviolet (UV) descriptions but flow to the same infrared (IR) theory. A well-known example in (2+1)D is the particle-vortex duality, which states that the XY model is dual to the Abelian Higgs model \cite{karch2016particle,peskindual,halperindual}. Recent developments further extend this understanding and discover many theories and their dual partners, altogether they form a web of duality\cite{seiberg_a-duality_2016}.

\begin{figure}[t]
\begin{center}
\includegraphics[width=\columnwidth]{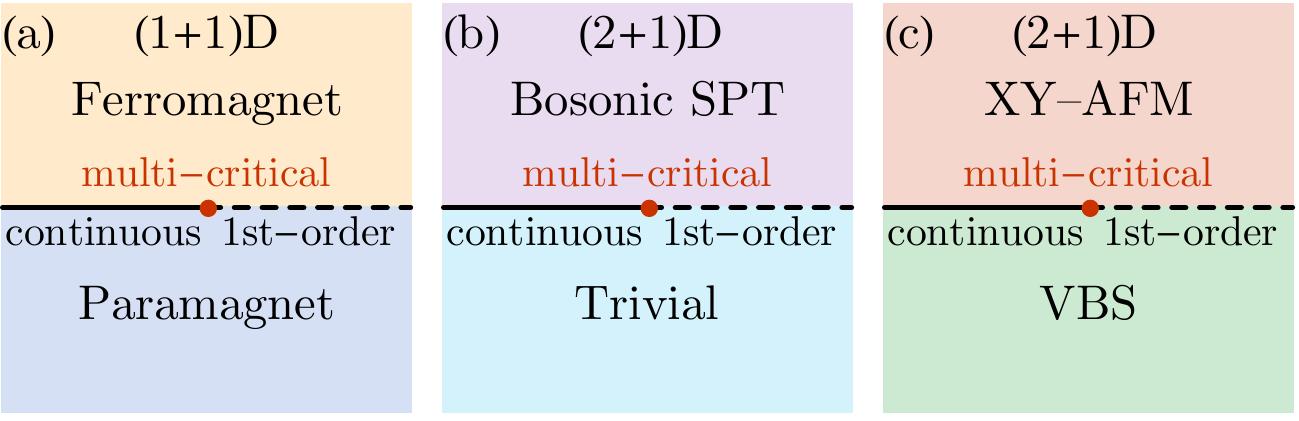}
\caption{Quantum phases related by the duality or emergent symmetry: (a) ferromagnetic (ordered) and the paramagnetic (disordered) phases across the Ising transition are related by the Kramers-Wannier duality, (b) bosonic symmetry protected topological (SPT) and trivial phases are related by the fermionic particle-vortex duality, (c) XY antiferromagnetic (AFM) and valence bond solid (VBS) phases are related by an emergent $\Z_2$ symmetry. In all phase diagrams, the vertical axis is the relevant perturbation that drives the transition between the duality/symmetry-related phases, and the horizontal axis is always taken to be the square of the transition-driving perturbation. In phase diagrams (b,c), the existence of continuous transitions between the adjacent phases is assumed, which corresponds to the $N_f=2$ QED$_3$ field theory without four-fermion interactions.}
\label{fig:dual}
\end{center}
\end{figure}

If the theory remains the same under a duality, the duality will be called a self-duality. For example, the Kramers-Wannier duality is a self-duality for the (1+1)D Ising model at the critical point. Recent studies \cite{chang_topological_2019,ji_categorical_2019,thorngren_fusion_2019,lichtman_bulk_2020} further propose to interpret the self-duality as a categorical symmetry, making connections to the fusion category of anyon excitations in the corresponding bulk topological order in one higher dimension. When the self-duality is imposed as a symmetry, the system is enforced to stay on the phase boundary between the two duality-related phases, leading to the self-duality protected criticality and multi-criticality\cite{aasen_topological_2016,buican_anyonic_2017,bal_mapping_2018,ji_noninvertible_2019,kong_a-mathematical_2020,chen_topological_2020}. For example, as illustrated in \figref{fig:dual}(a), in the presence of the Kramers-Wannier duality (enforcing $J=h$), a generic Ising chain (with all additional duality-allowed terms like $-K(X_{i}X_{i+1}+Z_{i-1}Z_{i+1})$) can either preserve the self-duality and remain gapless along the Ising critical line ($K<K_c$), or spontaneously break the self-duality and becomes gapped along the first-order transition line ($K>K_c$). The continuous and first-order Ising transitions are separated by a multi-critical point ($K=K_c$), i.e.~the tricritical Ising point\cite{blume_ising_1971,nienhuis_first-_1979}. The multi-critical point can be circumvented if the self-duality is explicitly broken (e.g.~by $J\neq h$). In this sense, the multi-criticality is protected by self-duality.

Similar continuous to first-order transition also happens in higher dimensions between the duality-related quantum phases. Here we will explore the (2+1)D example of self-duality-protected multi-criticality. In particular, we will consider the topological transition between the bosonic symmetry protected topological (SPT) phase and the trivial phase, as illustrated in \figref{fig:dual}(b), where the two phases across the transition are related by the self-duality\cite{xu_self-dual_2015,seiberg_a-duality_2016,hsin_levelrank_2016,cheng_series_2016,benini_comments_2017} of the quantum electrodynamics in (2+1)D (QED$_3$) with fermionic matters at flavor number $N_f=2$. This theory also describes the deconfined quantum critical point (DQCP)\cite{senthil_deconfined_2004,senthil_quantum_2004,levin_deconfined_2004} between the XY antiferromagnet (AFM) and the valence bond solid (VBS) in square-lattice quantum magnets with the easy-plane spin anisotropy, as shown in \figref{fig:dual}(c). In this case, the two phases are related by a $\Z_2$ subgroup of the emergent $\Oo(4)$ symmetry that maps the two-component XY-AFM order parameter to the two-component VBS order parameter. Imposing these emergent symmetries (including the self-duality) essentially promotes the tuning parameters to the fluctuating scalar fields and prohibits the explicit mass terms. This leads to a unified field theory that describes the continuous-to-first-order transition in these systems. Such multi-critical point lies in the universality class described by the QED$_3$-Gross-Neveu\cite{gracey_critical_1992,seiberg_a-duality_2016,karch2016particle} (QED$_3$-GN) theory. More generally, the Chern-Simons (CS) term for the gauge fields can be included to describe the multi-critical point of the exotic quantum phase transitions.

To further investigate the stability of this (2+1)D self-duality-protected multi-criticality, we extend the field theory to the large fermion flavor number (large $N_f$) limit, and use the $1/N_f$ expansion\cite{gracey_critical_1992,xu_square-lattice_2008,xu_renormalization_2008,jian_emergent_2017,you_from_2018,alanne_abelian_2018,boyack_deconfined_2019} to analyze the renormalization group (RG) flow of the fermion quartic operators, including the mass-mass $(\bar{\psi}M\psi)^2$ and current-current $(\bar{\psi}\gamma^\mu M\psi)^2$ interactions, at the QED$_3$-GN fixed point. Our analysis indicates that the DQCP and the multi-criticality can be driven to the first-order transition by current-current interactions. Such current-current interactions can be realized in the lattice spin model as a staggered dimer-dimer interaction (or stagger-$Q$) as proposed and observed in the recent quantum Monte Carlo (QMC) studies\cite{zhao_multicritical_2020,zhao_tunable_2020}. Unlike the conventional dimer-dimer interaction that couples the dimers along the vertical or horizontal directions on the square lattice, the stagger-$Q$ interaction couples the dimers along the diagonal direction. The QMC results indicate that such a stagger-$Q$ interaction may be responsible for driving the DQCP between continuous and first-order transitions (see \secref{sec:QMC} for more concrete discussion of the QMC results and our theoretical explanation).

The RG analysis can be further generalized to the QED$_3$-GN theory with additional Chern-Simons (CS) terms for the gauge field. Although there is a lack of known examples of self-dual theory with a non-zero-level CS term, a similar multi-critical point separating the continuous and first-order transition still exists and can be analyzed. The result can be applied to the direct transition between bosonic fraction quantum Hall (FQH) and superfluid (SF) phases in interacting boson systems\cite{barkeshli_continuous_2014,barkeshli_continuous_2015}.


\section{Self-Duality of $N_f=2$ QED$_3$}\label{sec:self_dual}

The fermionic particle-vortex duality\cite{son2015composite,metlitski2016particle} dualizes a free Dirac fermion theory to $N_f=1$ QED$_3$ theory with CS terms and the fermion operator is mapped to the fermion operator combined with gauge fluxes. Since CS terms break parity symmetry, the orientation reversed version of the fermionic particle-vortex duality is obtained by changing the sign of the CS terms. By combining the fermion particle-vortex duality and its orientation reversed version, one can obtain a duality between two $N_f=2$ QED$_3$ theories\cite{xu_self-dual_2015,benini_comments_2017,hsin_levelrank_2016} described by the following Lagrangians,



\begin{widetext}
\begin{align}
  &\ii\bar{\psi}_1 \idty{\slashed{D}}_{a+X} \psi_1+\ii\bar{\psi}_2 \idty{\slashed{D}}_{a-X} \psi_2 +\fpic{1}(a+Y)d(a+Y)+\fpic{2}(XdX-YdY) \label{eq:selfdual_1}\\
   & \Longleftrightarrow \ii\bar{\chi}_1 \idty{\slashed{D}}_{\ta+Y} \chi_1+ \ii\bar{\chi}_2\idty{\slashed{D}}_{\ta-Y} \chi_2+\fpic{1}(\ta+X)d(\ta+X),\label{eq:selfdual_2}
\end{align}
\end{widetext}
where $\psi_i,\chi_i$ are fermion fields, $\idty{\slashed{D}}_a \equiv \gamma^\mu(\idty{\partial}_\mu-\ii a_\mu)$ is the Dirac operator coupled to the $\U(1)$ gauge field $a$. $ada\equiv \epsilon_{\mu\nu\rho}a_\mu\partial_\nu a_\rho$ is understood as the exterior product $a\wedge da$, and the same applies for other CS terms. We adopt the convention as the lower case letters $a,\ta$ represent the dynamical $\U(1)$ gauge fields which will be integrated over in the path integral, and the upper case letters $X,Y$ represent the background gauge fields which are used to keep track of the $\U(1)_X$ and $\U(1)_Y$ global symmetries.

The two theories (at least) have the common UV symmetry $\U(1)_X\times \U(1)_Y$. For the $\gU(1)$ gauge theories in 2+1d, they automatically have an emergent global $\gU(1)_M$ magnetic symmetry due to the Bianchi identity $\epsilon^{\mu\nu\lambda}\partial_\mu F_{\nu\lambda} = 0$ where $F_{\nu\la}$ is the gauge field strength. The charged operator of this $\gU(1)_M$ symmetry is the magnetic monopole operator which creates the gauge flux and its coupling with the background gauge field are $\tpi adY, \tpi \ta d X$ in the both hand sides respectively. The symmetry charges of the operators are,
\begin{equation}
  \begin{array}{c|ccc}
       & \gU(1)_a & \gU(1)_X & \gU(1)_Y \\ \hline
    \cM_a & 1    & 0    & 1 \\ \hline
    \psi_1& 1    & 1    & 0 \\ \hline
    \psi_2& 1    & -1    & 0
  \end{array} \leftrightarrow
  \begin{array}{c|ccc}
       & \gU(1)_\ta & \gU(1)_X & \gU(1)_Y \\ \hline
  \cM_\ta  & 1     & 1    & 0 \\ \hline
    \chi_1& 1     & 0    & 1 \\ \hline
    \chi_2& 1     & 0    & -1
  \end{array}
\end{equation}
and the gauge invariant operators are built from these operators.

Renaming the fermion fields $\psi\leftrightarrow \chi$ will exchange $X\leftrightarrow Y$ and add a background term $\frac{2}{4\pi}(XdX-YdY)$ to the Lagrangian, the left-hand-side (LHS) \eqnref{eq:selfdual_1} and the right-hand-side (RHS) of \eqnref{eq:selfdual_2} of the duality will be swapped, therefore, establishes the self-duality.

This self-duality can also be understood as exchanging the ``electric charge'' and the ``magnetic charge''. On the LHS of the duality, the fermion field $\psi_i$ is charged under the $\U(1)_X$ flavor symmetry, and the magnetic monopole operator $\mathcal{M}_a$ which creates $2\pi$-flux for $a$ is charged under the magnetic $\U(1)_Y$ due to the mixed CS term $\tpi adY$ (note that $\cM_a$ is the bare magnetic monopole operator which is not gauge invariant due to the CS term $\frac{1}{4\pi}ada$, the gauge-invariant operators are the combination of the $\cM_a$ and fermion creation operators). However, on the RHS, the fermion field $\chi_i$ is charged under $\U(1)_Y$ and the magnetic monopole operator $\mathcal{M}_\ta$ is charged under $\U(1)_X$. This suggests that the fermion creation operators (resp. monopole operators) on the LHS become monopole operators (resp. fermion creation operators) on the RHS. More details of the self-duality are presented in \appref{sec:app_self-dual}

Here is a side-note on the conventions to regularize the fermion path integral:

One convention is that integrating out a single Dirac fermion in (2+1)D will contribute a $(-1)$-level CS term for the negative fermion mass and a $0$-level CS term for the positive fermion mass. Physically, fermions are doubled when putting on the lattice, one Dirac fermion is accompanied by a massive fermionic partner, otherwise, the single Dirac fermion will have parity anomaly in (2+1)D \citen{witten2016parity}. This convention assumes that the massive fermionic partner \textit{is not} integrated out beforehand and it is more explicit on the quantization of the level of Chern-Simons term, this is easier to analyze the symmetry charges of the operators since the magnetic monopole operator has charge $k$ if there is a level-$k$ CS term. We will use this convention in discussing the dualities of quantum field theories, such as the self-duality of \ntqed.

Another convention is that integrating out the fermion will contribute a $\frac{\mathrm{sgn}(m)}{2}$-level CS term, this assumes that the massive fermionic partner has been integrated out beforehand and this is relevant to the analysis of the scaling dimensions of the critical theory since the massive fermionic partner does not involve in the transition. Using the later convention, half level CS term will involve in the massless theory, and now the Chern-Simons level is effectively $-\frac{N_f}{2}+k$ where $N_f$ is the number of fermion flavors. We will adopt this convention in the discussion of renormalization group analysis on the critical behavior of the theory.

Schematically, the fermion theory with the level-$k$ CS term using the first convention is related to that using the second convention by,
\begin{equation}
  \underbrace{\ii \sum_{i=1}^{N_f}\bar{\psi}_i \idty{\slashed{D}}_a \psi_i+\fpic{k}ada}_\text{the 1st convention} \cong \underbrace{\ii \sum_{i=1}^{N_f}\bar{\psi}_i\slashed{D}_a \psi_i+\fpic{k-N_f/2}ada}_\text{the 2nd convention}.
\end{equation}
The duality presented in \eqnref{eq:selfdual_1} and \eqref{eq:selfdual_2} will be equivalent to the self-dual theory presented in \refcite{xu_self-dual_2015} by converting to the second convention of the fermion path integral regularization. However, both conventions have the same gauge-invariant operators and they yield the same response theories in the gapped phases. 

\subsection{Phase diagram}
The \ntqed has two relevant fermion mass deformations, the \emph{singlet mass} $\idm \bpsi \mathds{1} \psi \equiv \idm(\bar{\psi}_1\psi_1+\bar{\psi}_2\psi_2)$ and the \emph{triplet mass} $\trima \bpsi \sigma^3 \psi\equiv \trima(\bar{\psi}_1\psi_1-\bar{\psi}_2\psi_2)$, where $\sigma^i$ is the $i$-th Pauli matrix.
Under these mass deformations, one can integrate out the fermions and obtain the following effective theories for the background gauge fields \eqnref{eq:selfdual_1}\cite{xu_self-dual_2015,cheng_series_2016},
\begin{align}
  &\begin{cases}
    \fpic{2}(XdX-YdY) & \idm>0,\ \trima=0\\
    0 & \idm<0,\ \trima=0\\
  \end{cases} \label{eq:spt}\\
  &\begin{cases}
    \tpi ad(Y+X)+\frac{1}{4e^2}f^2+... & \trima>0\ \idm=0\\
    \tpi ad(Y-X)+\frac{1}{4e^2}f^2+... & \trima<0\ \idm=0\\
  \end{cases}.\label{eq:ssb}
\end{align}
where $e$ is the electron charge. The $...$ represents the gapped degrees of freedom that are not important at low energy since the low-energy physics is dominated by the Maxwell term $\frac{1}{4e^2}f^2$ and the first term which describes the gapless Goldstone boson associated to the broken symmetry $\gU(1)_{Y+X}$ or $\gU(1)_{Y-X}$.

\begin{figure}
  \centering
  \includegraphics[width=\columnwidth]{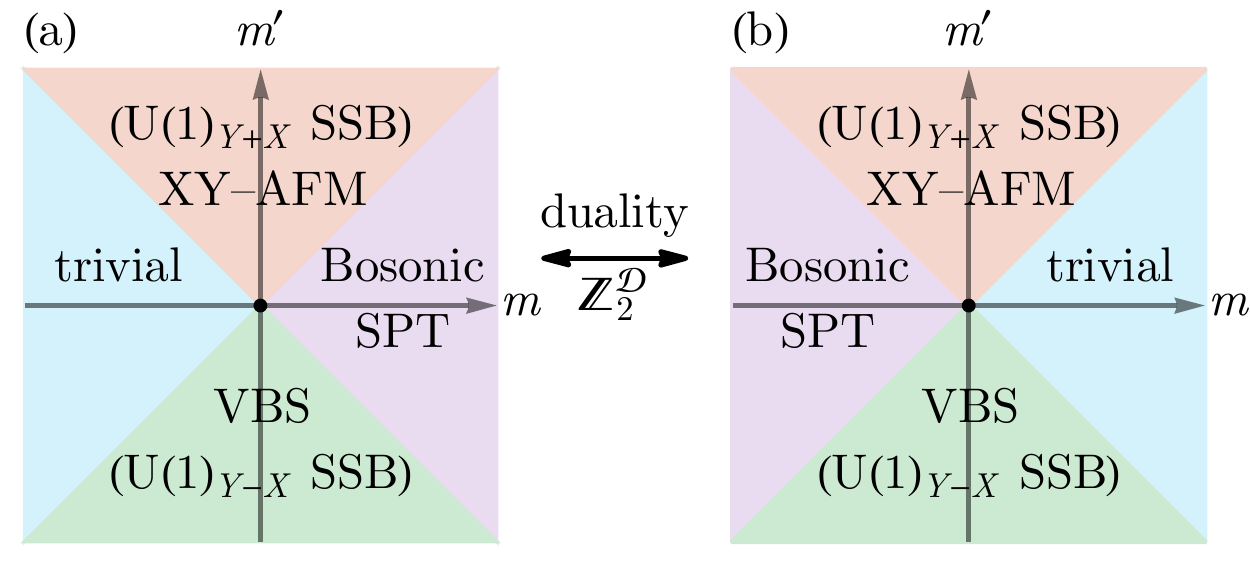}
  \caption{The phase diagram of \ntqed theory. The singlet mass $\idm$ drives the SPT transition between two symmetric phases, and the triplet mass $\trima$ drives AFM-VBS transition between two symmetry broken phases.}
  \label{fig:phase_diagram_quar}
\end{figure}

When the singlet mass $\idm$ is non-zero, the two response theories in \eqnref{eq:spt} differ by a $\U(1)_{X,2}\times\U(1)_{Y,-2}$ CS term, where the number indicates the level of the CS term, i.e.~$\fpic{2}(XdX-YdY)$, which corresponds to the topological response of a bosonic SPT state with $\U(1)_X\times\U(1)_Y$ symmetry\footnote{Since the gauge-invariant operators in UV are all bosonic (no single fermion operators), the resulting gapped phases can possibly connect to the bosonic theory.}. Therefore, the $\idm>0$ and $\idm<0$ phases should be ascribed to the topological and trivial SPT phases respectively\footnote{Which phase is topological/trivial is only a matter of convention, as the notion of SPT phases is only relative.}. When the triplet mass term $\trima$ is non-zero, the effective theories in \eqnref{eq:ssb} describe the Goldstone modes in the spontaneous symmetry breaking (SSB) phases with broken symmetries associated to $Y+X$ and $Y-X$ respectively (two different combinations of the generators of $\U(1)_X,\U(1)_Y$). In the context of square-lattice easy-plane quantum magnets\cite{Qin2017Duality,Ma2018Dynamical}, we might interpret $\U(1)_{Y+X}$ as the in-plane spin rotation symmetry and $\U(1)_{Y-X}$ as the lattice rotation symmetry (ignoring the discrete nature of the actual $C_4$ rotation), then the $\trima>0$ and $\trima<0$ phases could be identified as the XY-AFM and the VBS phases respectively. \figref{fig:phase_diagram_quar}(a) shows the phase diagram summarizing the above interpretations. Under the duality transformation, the singlet mass is odd ($\idm\to-\idm$) while the triplet mass is even ($\trima\to \trima$), which effectively swap the SPT and trivial phases but leaving the AFM and VBS phases unchanged (see \figref{fig:phase_diagram_quar}). To restore the original phase diagram after the duality transformation, one should exchange $\U(1)_X\leftrightarrow \U(1)_Y$ and add a background $\U(1)_{X,2}\times\U(1)_{Y,-2}$ CS term to the Lagrangian.

\subsection{Self-duality as a symmetry}
As pointed out in \refcite{benini_comments_2017,hsin_levelrank_2016}, the explicit UV symmetry $\U(1)_X\times \U(1)_Y$ in \eqnref{eq:selfdual_1} and \eqref{eq:selfdual_2} can be enhanced to the emergent symmetry $\frac{\SU(2)_X\times \SU(2)_Y}{\Z_2}\cong \SO(4)$ in the IR. Together with the self-duality $\Z_2^{\cD}$ which exchanges $\SU(2)_X\leftrightarrow\SU(2)_Y$ and attaches a $\SU(2)_{X,1}\times \SU(2)_{Y,-1}$ CS term (which falls back to the $\U(1)_{X,2}\times\U(1)_{Y,-2}$ CS term in the UV), the IR symmetry becomes $\SO(4)\rtimes \Z_2^{\cD}\cong \Oo(4)$. However, as the IR theory is shifted by the $\SU(2)_{X,1}\times \SU(2)_{Y,-1}$ background response under the self-duality transformation, the $\Z_2^\cD$ and the $\SO(4)$ have the mixed 't Hooft anomaly, thus they cannot be simultaneously coupled to the background gauge fields and promoted to the dynamical ones. Nonetheless, it can be viewed as the boundary of a (3+1)D SPT with the full $\Oo(4)$ symmetry. With appropriate counterterm in the bulk, the whole system can also have time-reversal symmetry $\Z_2^\mathsf{T}$, altogether gives $\Oo(4) \times \Z_2^\mathsf{T}$ as suggested in \refcite{wang_deconfined_2017}.

Note that the singlet mass $\idm$ is invariant under $\SO(4)$ but is odd under $\Z_2^\cD$, while the triplet mass $\trima$ explicitly breaks $\SO(4)$ (as it is in the $(\mathbf{3,3})$ representation\cite{cheng_series_2016,wang_deconfined_2017} of $\SU(2)_X\times \SU(2)_Y$) but is even under $\Z_2^\cD$. Hence, if both the emergent $\SO(4)$ and the self-duality $\Z_2^\cD$ symmetries are imposed, no fermion bilinear mass could be included in the Lagrangian.

\subsection{Self-duality protected multi-criticality}\label{section:critical_behavior}
Although the mass term cannot be added to the Lagrangian, squares of the mass term still can, which may take the form of four-fermion interactions $(\bpsi M^a \psi)^2$, where $M^a$s are mass matrices acting on the flavor indices. Adding these mass-squared deformations to the QED theory \eqnref{eq:selfdual_1} could potentially drive the theory to new fixed points\cite{jian_emergent_2017}. The fate of the self-duality $\Z_2^\cD$ and the $\SO(4)$ symmetry depends on the RG flow of such mass-squared deformations. If both symmetries are preserved, the theory will remain critical (as no mass deformation is allowed), which describes the continuous transition between AFM and VBS phases (as well as the transition between SPT and trivial phases), which is also known as the $\Oo(4)$ DQCP. When the self-duality $\Z_2^\cD$ symmetry is spontaneously broken, the SPT transition becomes first-order. When the emergent $\SO(4)$ symmetry (more specifically the $\Z_2$ subgroup that swaps $\U(1)_{Y+X}$ and $U(1)_{Y-X}$) is spontaneously broken, the AFM-VBS transition becomes first-order. These first-order transitions are separated from the continuous transition by the multi-critical points/lines. We will analyze the RG flow of the generic four-fermion interactions at these multi-critical points, aiming to understand how certain kinds of interactions can drive the DQCP from a continuous transition to a first-order transition.

The multi-critical point happens when Dirac fermion masses change the sign. To analyze the scaling dimensions of the operators at the multi-critical point, we do not need to include the massive fermionic parton which is served to cancel the subtlety in the fermion path integral regularization. We rewrite \eqnref{eq:selfdual_1} as
\begin{equation}
  \ii \bar{\psi}_1 \slashed{D}_{a+X} \psi_1+\ii \bar{\psi}_2 \slashed{D}_{a-X} \psi_2 +\tpi adY+\fpi(XdX-YdY).
\end{equation}
The CS terms look different from \eqnref{eq:selfdual_1}, because we integrate out the massive fermionic partners beforehand and it corresponds to the second convention as discussed in the last three paragraphs of Sec.~\ref{sec:self_dual}, following from \refcite{xu_self-dual_2015}. Note that the changing of convention will not change the gauge invariant operators as well as the different gapped phases. The background gauge fields $X$ and $Y$ won't affect the dynamics and can be set to zero. Adding the mass-squared deformations amounts to promoting the mass terms $\idm$ and $\trima$ to the dynamic scalar fields $\phi_1$ and $\phi_2$, that are coupled to the fermions via Yukawa-type couplings $\phi_a \bpsi M^a \psi$, this can also be seen by using the Hubbard–Stratonovich transformation. Together with their own boson mass terms $r_a\phi_a^2$, the action reads as,
\begin{align}\label{eq:qed3_GNY_critical}
  &\sum_{i=1}^2\ii \bar{\psi}_i \slashed{D}_{a} \psi_i + \phi_1 \bar{\psi} \mathds{1} \psi+ \phi_2 \bar{\psi} \sigma^3 \psi \nonumber\\
  &+\sum_{a=1}^2\frac{1}{2g^2} \phi_a (r_a-\partial^2)\phi_a +\frac{\lambda}{4} (\phi_a \phi_a)^2.
\end{align}
For each scalar field $\phi_a$, the boson mass $r_a$ has a corresponding critical value $r_{a,c}$. When $r_a \gg r_{a,c}$, the boson is gapped and $\vv{\phi_a}=0$. When $r_a \ll r_{a,c}$, the boson is condensed, such that $\vv{\phi_a}\neq 0$ and the symmetry is spontaneously broken. This will dynamically generate the corresponding fermion mass terms. We may loosely set $r_{a,c}=0$ and assume the bosons are critical when $r_a=0$ in the following discussion.

\begin{figure}
  \centering
  \includegraphics[width=0.6\columnwidth]{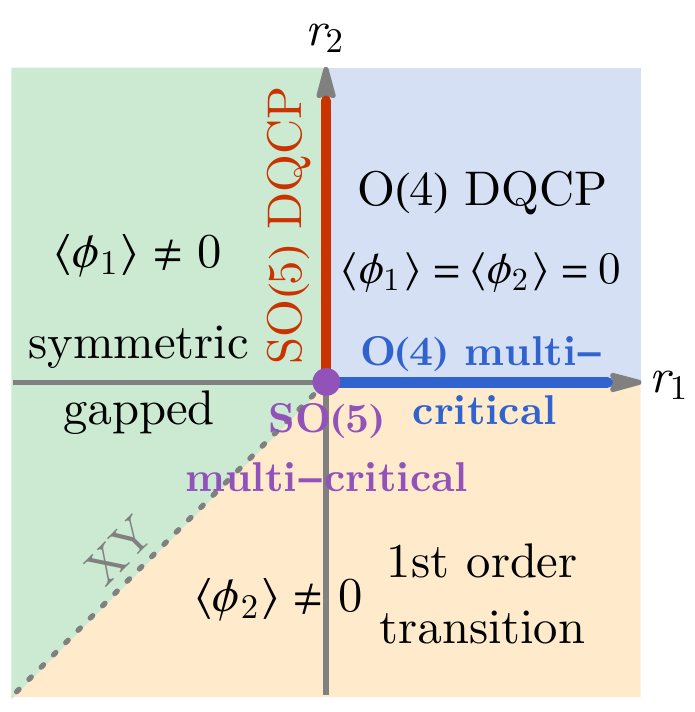}
  \caption{Mean-field phase diagram of \eqnref{eq:qed3_GNY_critical}.}
  \label{fig:phase_diagram_multi}
\end{figure}

The qualitative phase diagram of \eqnref{eq:qed3_GNY_critical} is shown in \figref{fig:phase_diagram_multi}, {which can be considered as the extension of the \figref{fig:phase_diagram_quar}'s origin, since no fermion mass terms $m,m'$ are added in the \eqnref{eq:qed3_GNY_critical}}. In the phase diagram, when $r_1,r_2\gg0$ (the blue region), both bosons are gapped, leaving \eqnref{eq:qed3_GNY_critical} to be the $N_f=2$ QED$_3$ theory at low energy. As discussed previously, this theory has an emergent $\Oo(4)$ symmetry and describes the continuous DQCP transition between the AFM and VBS phases (i.e.~between the $\U(1)_{Y+X}$ and $\U(1)_{Y-X}$ SSB phases) when tuning the triplet fermion mass $\trima$ externally. If $r_1$ is at its critical value and $r_2\gg 0$ (along the red line), the critical theory becomes $N_f=2$ QED$_3$-Gross-Neveu model, which describes the continuous DQCP with emergent $\SO(5)$ symmetry as proposed in \refcite{wang_deconfined_2017}. If instead, $r_2$ is at its critical value and $r_1\gg 0$ (across the blue line), the theory describes the multi-criticality between the $\Oo(4)$ DQCP and the first-order AFM-VBS transition. If both $r_1$ and $r_2$ are critical (the purple point), the theory describes the multi-criticality between the $\SO(5)$ DQCP and the first-order AFM-VBS transition.

To see that the $\phi_2$ condensed phase (the orange region) corresponds to the first-order AFM-VBS transition, we consider driving the AFM-VBS transition by an external triplet mass $\trima$. The actual mass term seen by the fermion will be $(\trima+\vv{\phi_2})\bpsi \sigma^3 \psi$, meaning that the driving parameter $\trima$ needs to overcome the expectation value $\vv{\phi_2}$ in order to change the sign of the triplet mass effectively and switch the system from one phase to another. Therefore $\vv{\bpsi \sigma^3 \psi}$ will exhibit the hysteresis behavior as $\trima$ is tuned back and forth, which manifests the first-order transition. Without the external driving ($\trima=0$), the ground state will be degenerated between AFM and VBS phases.

On the other hand, the $\phi_1$ condensed phase (the green region) is a symmetric gapped phase whose ground state is degenerated between topological and trivial SPT phases, which may as well be interpreted as the 1st-order SPT transition if the singlet mass $\idm$ is tuned externally. The $\phi_1$ condensed phase and the $\phi_2$ condensed phase do not coexist, because they compete with each other to gap out the fermion, and the ground state is determined by the condensate that has a larger vacuum expectation value $|\vv{\phi_a}|$. When the competition reaches a balance at $|\vv{\phi_1}|=|\vv{\phi_2}|$ (along the gray dashed line), it triggers a direct transition between the symmetric and the SSB phases (either the in-plane magnetic order or the VBS order), which is of the 3d XY universality.

The multi-criticality between the continuous and first-order transitions cannot be circumvented in the presence of the anomalous $\Oo(4)$ symmetry. However, it is possible that the protecting symmetry may be broken spontaneously under other potentially relevant perturbations, such that the $\Oo(4)$ DQCP is not stable in general. In the following, we will explore this possibility by analyzing the effect of generic four-fermion interactions in the QED-GN theory using the large-$N$ renormalization group (RG) approach.

\section{Large-$N$ Renormalization Group Analysis}\label{sec:model}

\subsection{QED-Gross-Neveu-Chern-Simons theory}

We extend \eqnref{eq:qed3_GNY_critical} to $N_f$ flavors of Dirac fermions $\psi=(\psi_1,\cdots,\psi_{N_f})^\intercal$ coupled to the dynamical $\U(1)$ gauge field, together with Yukawa-type couplings to $N_b$ flavors of scalar bosons $\phi_a$ ($a=1,...,N_b$). The bosons will have their kinetic terms and can be tuned critical by the $r_a$ parameters. We also add the level-$k$ CS term for the dynamical $\U(1)$ gauge field (to be general) and consider the QED$_3$-Gross-Neveu-Chern-Simons (QED-GN-CS) theory as follows
\begin{equation}\label{eq:general_model_main}
\begin{split}
&\mathcal{L}=\bar{\psi} (\mathds{1}_{N_f}\otimes \gamma^{\mu})(\partial_\mu -\mathsf{i}a_\mu ) \psi + \phi_a \bar{\psi} (M^a \otimes \mathds{1}_2) \psi\\
&+ \frac{1}{2g^2} \phi_a (r_a-\partial^2)\phi_a +\frac{\lambda}{4} (\phi_a \phi_a)^2 \\
&+ \frac{\mathsf{i}k}{4\pi} \epsilon^{\mu \nu \lambda} a_\mu \partial_\nu a_\lambda+\frac{1}{4e^2}f_{\mu\nu}f^{\mu\nu}.\\
\end{split}
\end{equation}
Here, matrices $\mathds{1}_{N_f},M^a$ act on the flavor space, while matrices $\mathds{1}_2,\gamma^\mu$ act on the spinor space.
We take the $\ga$-matrices to be $(\si^3,\si^1,\si^2)$. $M^a$s are vertices of Yukawa couplings associated with fermion bilinear masses, which are assumed to be orthogonal to each other such that $\tr(M^a M^b)=\mathbb{M}\delta_{ab}$. The last term is the Maxwell term, with the gauge curvature defined as $f_{\mu\nu}=\partial_\mu a_\nu-\partial_\nu a_\mu$.

The multi-critical points/lines in the phase diagram \figref{fig:phase_diagram_multi} correspond to tuning one or more scalar bosons to critical. We assume that all scalar fields in the effective theory \eqnref{eq:general_model_main} correspond to the critical bosons (other gapped bosons will be dropped from the effective theory automatically). The theory is tuned to the QED-GN-CS fixed point. The boson mass term $(r_a-r_{a,c})\phi_a^2$ is a relevant perturbation that drives the system away from the multi-criticality. It also is possible that some types of fermion interactions may flow to the boson mass term $\phi_a^2$, as it is equivalent to the mass-mass interaction $(\bar{\psi} (M^a \otimes \mathds{1}_2) \psi)^2$ under the Hubbard–Stratonovich transform. Such fermion interactions will appear relevant at the QED-GN-CS fixed point and can drive the system away from multi-criticality as well.

\subsection{Renormalization of four-fermion interactions}\label{sec:scaling_4fermion}
To explore this possibility, we carry out a systematic study of the scaling dimension of four-fermion interactions at the QED-GN-CS fixed point (see \appref{appendix:largeN} for technical details). We will follow the large-$N_f$ expansion approach recently developed for the QED$_3$-GN model in \refcite{boyack_deconfined_2019}, where the scaling dimensions of fermion and boson bilinear operators were analyzed. Here, we will carry over the analysis to four-fermion operators, which has not been presented yet. To be more general, we also include a CS term, such that our result could potentially be applied to other DQCP such as the superfluid to bosonic fractional quantum Hall transition (described by the QED-GN-CS fixed point at level $k=1$\cite{barkeshli_continuous_2014}).

In particular, our scheme to extend \eqnref{eq:qed3_GNY_critical} to large $N_f$ corresponds to generalizing the fermion flavor symmetry group from $\SU(2)\rightarrow \SU(2N)$, such that the fermion flavor number scales as $N_f=2N$ with $N\to\infty$. The Yukawa vertices are generalized to
\begin{equation}
  \{M^a\}=\{\dsi_2,\sigma^3\}\rightarrow \{M^a_N\} = \{\dsi_2,\sigma^3\}\otimes \dsi_N.
\end{equation}
where $\{M^a\}$ denotes the set formed by $M^a$s, similar for $\{V^\al\}$. The perturbative interactions are,
\begin{equation}\label{eq:Lint}
  \mathcal{L}_\mathrm{int} = u_{\al,m}(\bar{\psi} V^\al \otimes \dsi_2 \psi)^2+u_{\al,\mu}(\bar{\psi} V^\al \otimes \ga^\mu \psi)^2
\end{equation}
where $V^\al = \sigma^\al\otimes \dsi_N$ ($\alpha=0,1,2,3$). $u_{\al,m}$, $u_{\al,\mu}$ represent the coupling coefficient of the mass-mass interactions and the current-current interactions respectively, which can be combined to a vector $u_{\al,i}=(u_{\al,m}, u_{\al,0},u_{\al,1},u_{\al,2})^\intercal$ in each $\alpha$-channel.
The RG equations for $u_{\al,i}$ takes the following general form,
\begin{equation}
  \frac{d u_{\al,i}}{d \ell} =\left(-1+\frac{64}{3\pi^2 N_f} \mM_{(\al,i),(\beta,j)}\right)u_{\beta,j}
\end{equation}
where the repeated indices are summed over and $\mM$ is a matrix with entries given by the $\mathcal{O}(1/N_f)$ corrections, the detailed calculations are presented in \appref{appendix:largeN}. One can further diagonalize $\mM$ to find the eigen-channels.
We take $N_f\rightarrow 2$ to restore the case of \eqnref{eq:qed3_GNY_critical}. {The large-$N_f$ analysis is not well controlled for small $N_f$, as sub-leading corrections may not be sufficiently small. However, in our case, we assume the \ntqed has the IR conformal fixed point which is suggested by the QMC simulation\citen{Qin2017Duality} and then perform the analysis on the perturbative four-fermion interactions. It turns out that our large-$N_f$ RG results are consistent with the latest QMC simulation\citen{zhao_multicritical_2020,zhao_tunable_2020,yang2021quantum}.}

\noindent\textbf{The first quadrant, $\Oo(4)$ DQCP}: Without the contribution from the critical bosons, there is no relevant channel for $\alpha=0$. But for $\al=1,2,3$, it has one relevant channel,
\begin{equation}\label{eq:o4_4fermion}
  \frac{d u_{\al,i}}{d \ell} = 2.24 u_{\al,i},\quad \mathrm{with}\  u_{\al,i}=(3,1,1,1)^\intercal,
\end{equation}
and the spatio-temporal anisotropic channels are irrelevant.
Therefore the mass-mass interaction can be generated from the current-current interaction under the RG flow, which could potentially drive the $\Oo(4)$ DQCP to a first-order transition (if the generated mass-squared interaction is strong enough to overcome the bare $r_2$ term).

{
With large-$N_f$, $u_{\alpha,i}$ are independent parameters. But for $N_f = 2$ (i.e. $N=1$), the Fierz identity demands the uniform combination $\sum_{\alpha = 1,2,3} u_{\alpha,i}$ ``fuses'' into the $\alpha = 0$ channel, which is irrelevant. Additionally, the explicit $\U(1)_{X}\times\U(1)_{Y}$ symmetry guarantees $u_{1,i} = u_{2,i}$, hence for $N_f=2$, there is only one independent channel of the relevant four-fermion interaction with $\alpha = 3$.
}

\noindent\textbf{The positive-$r_2$ axis, $\SO(5)$ DQCP}: In this case, the scalar boson associated to the singlet mass is critical, $\{M^a\}=\dsi_2$. There is still no relevant channel for $\alpha=0$. For $\al=1,2,3$, it has the same relevant channel as the previous case,
\begin{equation}\label{eq:so5_4fermion}
  \frac{d u_{\al,i}}{d \ell} = 1.70 u_{\al,i},\quad \mathrm{with}\  u_{\al,i}=(3,1,1,1)^\intercal.
\end{equation}
Hence, the stagger-Q term still overlaps with the relevant channel at $\SO(5)$ DQCP fixed point. Similarly, as discussed in the last paragraph, for $N_f = 2$, there is only one independent channel of the relevant four-fermion interaction with $\alpha = 3$.

\noindent\textbf{The positive-$r_1$ axis and the origin}: Both cases are more involved. The positive-$r_1$ axis describes the transition between the $\Oo(4)$ DQCP and first-order transition, and the origin is a multi-critical point where 3 critical lines joins. Both $\phi_1$ and $\phi_2$ scalar fields are critical at the origin, such that the Yukawa vertices are $\{M^a\}=\{\dsi_2,\sigma^3\}$. The eigen-channels will have mixture of $V^0,V^3$ or $V^1,V^2$, because $M^a$ will mix $V^0$ with $V^3$ as well as $V^1$ with $V^2$. Considering $\{V^\al\} =\{V^0,V^3\}$, there is one relevant channel with $u_{03}\equiv(u_{0,i};u_{3,i})=(-0.03,-0.071,-0.071,-0.071;0.82,0.32,0.32,0.32)^\intercal$, and the RG equation reads
\begin{equation}
\begin{split}
  \frac{d u_{03}}{d \ell} &= 1.89 u_{03}\quad\text{(positive-$r_1$ axis)},\\
  \frac{d u_{03}}{d \ell} &= 1.35 u_{03}\quad\text{(origin)}.
\end{split}
\end{equation}
The detailed calculation is presented in \appref{appendix:largeN}. With one more critical boson at the origin compared to the positive-$r_1$ axis, the RG eigenvalue of the relevant interaction is smaller at the $\SO(5)$ multi-critical point compared to the $\Oo(4)$ multi-critical line.



\section{Implications of RG Analysis}
\subsection{Consequence of the relevant interactions}

\begin{figure}[htbp]
\begin{center}
\includegraphics[width=0.6\columnwidth]{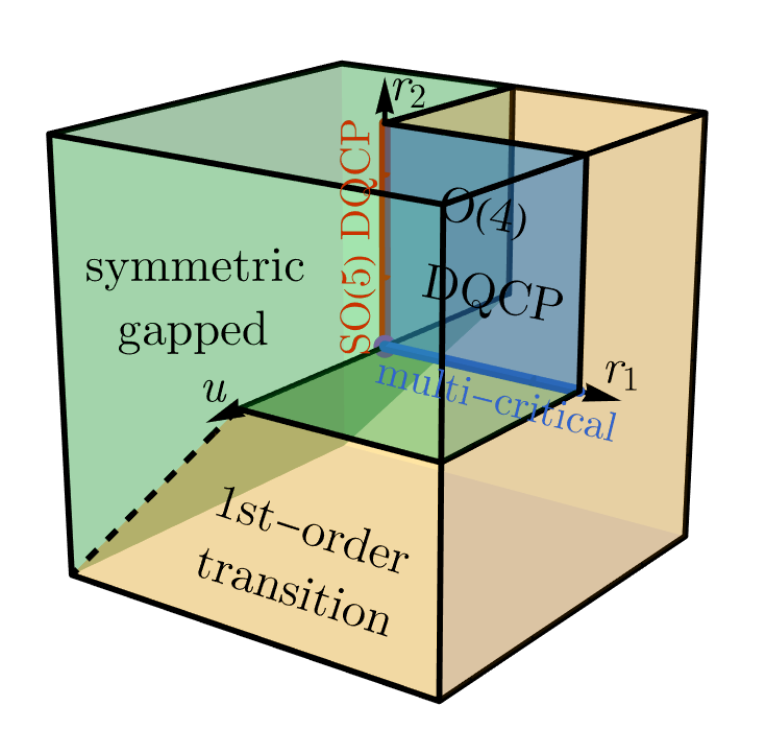}
\caption{Extended phase diagram in the presence of relevant interaction $u$. The $u=0$ plane corresponds to the phase diagram in \figref{fig:phase_diagram_multi}.}
\label{fig:phase 3D}
\end{center}
\end{figure}

The RG analysis suggests that the $\SO(5)$ and $\Oo(4)$ DQCP may not be stable against the perturbation of certain Lorentz symmetry breaking four-fermion interactions in the field theory. The interaction is relevant and flows to the following form
\begin{equation}\label{eq:relevant interaction}
\mathcal{L}_\text{int}=u(3(\bar{\psi}\sigma^3\psi)^2+(\bar{\psi}\sigma^3\gamma^\mu\psi)^2).
\end{equation}
Depending on the sign of the coefficient $u$, the interaction may drive different instabilities of the QED theory. By analyzing all possible Wick decomposition of the interaction term, we found the leading eigen decompositions with both positive and negative interaction strength is $\mathcal{L}_\text{int}=u(\bar{\psi}\sigma^3\psi)^2+\cdots-u(\bar{\psi}\psi)^2$. Therefore, if $u<0$, the interaction favors the condensation of the triplet mass term $\bar{\psi}\sigma^3\psi$, or equivalently the scalar field $\phi_2$ that couples to it. In this case, the emergent $\SO(4)$ symmetry is spontaneously broken, and the AFM-VBS transition becomes first-order. On the other hand, if $u>0$, the interaction favors the condensation of the singlet mass term $\bar{\psi}\psi$, or equivalently the corresponding scalar field $\phi_1$, which spontaneously breaks the self-duality and results in the symmetric gapped state. \figref{fig:phase 3D} shows the extension of the phase diagram in the presence of four-fermion interaction.

The next leading eigen decompositions of the interaction are the singlet pairing channels $-\frac{2}{3}u|\psi^\intercal \si^2\ga^0\ga^x \psi|^2$ and $-\frac{2}{3}u|\psi^\intercal \si^2\ga^0\ga^y \psi|^2$ with slightly less interaction strength. When $u>0$, the system may condense the Cooper pairs $\psi^\intercal \si^2\ga^0\ga^{x,y} \psi$, breaking the Lorentz symmetry. Since this term commutes with some of the kinetic terms in the Hamiltonian, it will split the Dirac points in the momentum space but will not gap out the fermions. It will also Higgs the $\U(1)$ gauge group down to $\Z_2$. Therefore, it opens the possibility for the gapless $\Z_2$ spin liquid phase instead of the symmetric gapped phase away from the multicritical point, which provides a candidate scenario for the phase diagram observed in the recent QMC study \refcite{yang2021quantum} where the first-order transition and the gapless $\Z_2$ spin liquid phase are separated by the multicritical point. Another scenario of the gapless $\Z_2$ spin liquid phase near the DQCP is recently proposed in \refcite{shackleton2021deconfined}. The Lorentz symmetry is also broken by the Higgs field. However, the fermion flavors are doubled in that proposal compared to ours, thus it describes a different gapless $\Z_2$ spin liquid phase (see \appref{sec:app_z2_spin_liquid} for details). For example, the entanglement entropy contributed from the massless degrees of freedom will be different, which could be distinguished in future numerical studies.

\subsection{Role of the stagger-$Q$ perturbation}\label{sec:QMC}

\begin{figure}[htbp]
\begin{center}
\includegraphics[width=0.64\columnwidth]{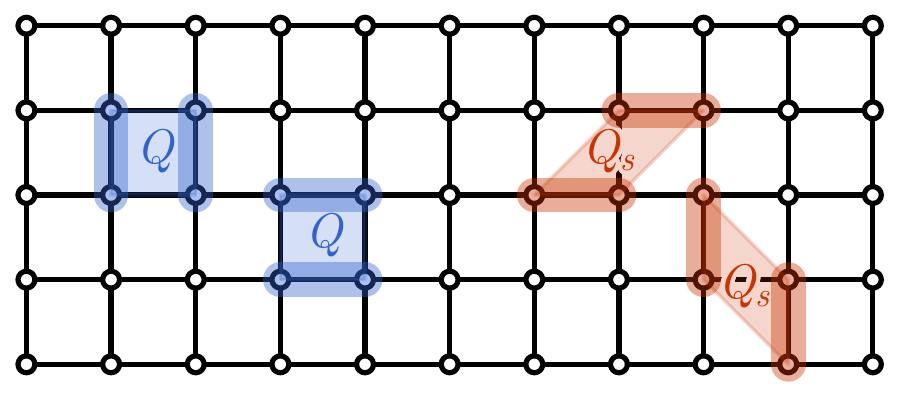}
\caption{Illustration of the (standard) $Q$ term (in blue) and the stagger-$Q$ term ($Q_s$, in red) on the square lattice. Both are dimer-dimer interactions, but along different directions.}
\label{fig:Qterm}
\end{center}
\end{figure}

Recent QMC studies revealed the possibility of tuning the DQCP between continuous and first-order transitions\cite{zhao_multicritical_2020,zhao_tunable_2020}.
In particular, the stagger-$Q$ term (denoted by $Q_s$, or the so-called $Z$-deformation) was proposed in \refcite{zhao_multicritical_2020} as a modification of the $J$-$Q$ model,
\eqs{\label{eq:Qs}
H&=H_{JQ}+H_{Q_s},\\
H_{JQ}&=-J\sum_{i}P_{i}^x-Q\sum_{i}P_{i}^xP_{i+\hat{y}}^x+(x\leftrightarrow y),\\
H_{Q_s}&=-Q_s\sum_{i}P_{i}^xP_{i+\hat{x}+\hat{y}}^x+(x\leftrightarrow y),}
where $P_{i}^{x}=1/4-\vect{S}_i\cdot\vect{S}_{i+\hat{x}}$ and $P_{i}^{y}=1/4-\vect{S}_i\cdot\vect{S}_{i+\hat{y}}$ are the dimer operators on the $x$ and $y$ bonds respectively. The stagger-$Q$ term $Q_s$ favors a staggered VBS pattern, and hence the name. The illustration of the $Q$ term and the stagger-$Q$ term is shown in Fig.~\ref{fig:Qterm}. Another version of the stagger-$Q$ term that involves three dimers interacting along the diagonal direction is studied in \refcite{zhao_tunable_2020}. The three-dimer stagger-$Q$ term has the same symmetry as the two-dimer stagger-$Q$ term, and shares the similar physical effect (both favors the same staggered VBS order). The QMC phase diagram in \refcite{zhao_tunable_2020} explicitly shows that the stagger-$Q$ term can drive the DQCP to a first-order transition. We will connect this observation to our field-theory analysis.

In the momentum space, the stagger-$Q$ term should correspond to the dimer-dimer interaction near momentum $(\pi,\pi)$, which can be argued as follows. Let $P_{\vect{q}}^{x,y}=\sum_{i}P_{i}^{x,y} e^{-\ii \vect{q}\cdot \vect{r}_i}$ be the dimer operator of momentum $\vect{q}$. A large $Q_s$ term favors the dimer to order in the staggered pattern (along the diagonal direction), which corresponds to the condensation of the dimer order parameter at momentum $\vect{q}=(\pi,\pi)$, i.e.~$\langle P_{(\pi,\pi)}^{x}\rangle\neq 0$ or $\langle P_{(\pi,\pi)}^{y}\rangle\neq 0$. Therefore, the effect of the stagger-$Q$ interaction $H_{Q_s}$ can be expressed as
\eq{\label{eq:HQs1}H_{Q_s}\sim -Q_s\big((P_{(\pi,\pi)}^x)^2+(P_{(\pi,\pi)}^y)^2\big),}
because a large $Q_s$ in \eqnref{eq:HQs1} also promotes the ordering of $P_{(\pi,\pi)}^{x,y}$, matching the effect of $H_{Q_s}$ in the real space \eqnref{eq:Qs}.

At low-energy, the dimer fluctuation near momentum $(\pi,\pi)$ should correspond to the spatial component of the Noether current associated with the emergent $\U(1)_{Y-X}$ symmetry that rotates the VBS order parameters:
\eq{\label{eq:Pj}P_{(\pi,\pi)}^x\sim j_\text{VBS}^y, P_{(\pi,\pi)}^y\sim j_\text{VBS}^x.}
This mapping was derived in \refcite{Wang2019Dynamics} from the fermionic parton construction. A simple symmetry argument is as follows. We first notice that $P_{(\pi,0)}^{x}$ and $P_{(0,\pi)}^{y}$ are the VBS order parameters favored by the standard $Q$ term in the $J$-$Q$ model. They can be combined into a complex order parameter $\Psi_\text{VBS}=P_{(\pi,0)}^{x}+\ii P_{(0,\pi)}^{y}$. The $\U(1)_{Y-X}$ rotation corresponds to $\Psi_\text{VBS}\to e^{\ii\theta}\Psi_\text{VBS}$, therefore the associated current operator should be
\eqs{j_\text{VBS}^x&=\ii\Psi^\dagger_\text{VBS}\partial_x\Psi_\text{VBS}+\text{h.c.}\\
&=P_{(0,\pi)}^{y}\partial_x P_{(\pi,0)}^{x}- P_{(\pi,0)}^{x}\partial_x P_{(0,\pi)}^{y},\\
j_\text{VBS}^y&=\ii\Psi^\dagger_\text{VBS}\partial_y\Psi_\text{VBS}+\text{h.c.}\\
&=P_{(0,\pi)}^{y}\partial_y P_{(\pi,0)}^{x}- P_{(\pi,0)}^{x}\partial_y P_{(0,\pi)}^{y}.\\}
Thus both $j_\text{VBS}^{x}$ and $j_\text{VBS}^y$ carry the total momentum $(\pi,\pi)$ (as a summation of $(\pi,0)$ and $(0,\pi)$). Under the (site-centered) reflection about the $y$ axis, i.e.~$(x,y)\to(-x,y)$, we have $(P^x,P^y)\to(-P^x,P^y)$, $(\partial_x,\partial_y)\to(-\partial_x,\partial_y)$, thus $(j_\text{VBS}^x,j_\text{VBS}^y)\to (j_\text{BVS}^x,-j_\text{BVS}^y)$ transforms as a pseudo-vector. Similarly, under the reflection $(x,y)\to(x,-y)$, we have $(j_\text{VBS}^x,j_\text{VBS}^y)\to (-j_\text{BVS}^x,j_\text{BVS}^y)$. Furthermore, $j_\text{VBS}^{x,y}$ does not transform under spin rotation symmetry. All these symmetry properties are precisely matched by \eqnref{eq:Pj}, which speaks for its validity.

Using the operator correspondence in \eqnref{eq:Pj}, \eqnref{eq:HQs1} can be casted into
\eq{\label{eq:HQs2}H_{Q_s}\sim -Q_s\big((j_\text{VBS}^y)^2+(j_\text{VBS}^x)^2\big),}
which identifies the stagger-$Q$ term to the current-current interaction in the spatial channel. We can make further connection to the field theory. Since the $\U(1)_{Y-X}$ symmetry is generated by $\psi_1^\dagger \psi_1 - \psi_2^\dagger \psi_2$ in the $N_f=2$ QED$_3$ theory, the corresponding Noether current should be $j_\text{VBS}^\mu=\bar{\psi}\sigma^3\gamma^\mu\psi$, therefore the current-current interaction in \eqnref{eq:HQs2} further translates to the four-fermion interaction in \eqnref{eq:Lint} with $u_{3,i}\propto Q_s (0,0,1,1)^\intercal$. According to the RG analysis above, the current-current interaction will generate the mass-mass interaction and flow towards the combined interaction in \eqnref{eq:relevant interaction}.

{Since the $u$ term in \eqnref{eq:relevant interaction} corresponds to the stagger-$Q$ term in the lattice model, the original $J$-$Q$ model may be very close to $u=0$, i.e. the QED-GN fixed point in the field theory, though $u$ should never be precisely zero. But the stagger-$Q$ term in the lattice model will turn on a non-negligible $u$ term in the field theory which is relevant at the QED-GN fixed point, therefore render the transition first order, as was observed numerically. In fact, according to \eqnref{eq:so5_4fermion}, our calculation of the scaling dimension of the relevant four fermion term is $1.3 = 3 - 1.7$ at the $\SO(5)$ DQCP, which is close to the observed scaling dimension of the stagger-$Q$ deformation of the $J$-$Q$ model ($\Delta_Z \sim 1.4$ in \refcite{zhao_multicritical_2020}). }

The above field theory understanding also applies to the easy-plane $J$-$Q$ model\cite{Qin2017Duality, Ma2018Dynamical},
\eqs{\label{eq:EPJQ}
H&=H_{JQ}+H_{\Delta},\\
H_{\Delta}&=-J\Delta\sum_{i}S_i^zS_{i+\hat{x}}^z+(x\leftrightarrow y),}
where the parameter $\Delta$ tunes the easy-plane anisotropy. $\Delta=0$ is the $\SU(2)$ isotropic limit, and $\Delta=1$ is the $\U(1)\rtimes\Z_2$ easy-plane limit.

\begin{figure}
  \centering
  \includegraphics[width=0.62\columnwidth]{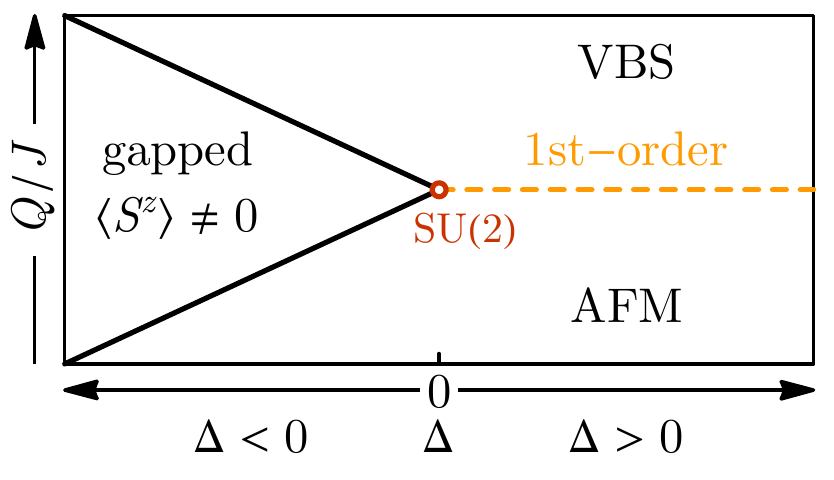}
  \caption{Schematic phase diagram of the easy-plane $J$-$Q$ model \eqnref{eq:EPJQ}.}
  \label{fig:easy_plane}
\end{figure}

Tuning $\Delta$ away from 0 breaks the spin $\SU(2)$ symmetry and {the $u$ term should in principle also exist for the easy-plane $J$-$Q$ model, but because it is more relevant compared to that in the $\SU(2)$ symmetric case (according to \eqnref{eq:o4_4fermion} and \eqnref{eq:so5_4fermion}), the easy-plane $J$-$Q$ model may be a first-order transition more obviously than the isotropic limit.} Based on the phase diagram \figref{fig:phase 3D}, the system will either enter an intermediate symmetric gapped phase or exhibit a first-order AFM-VBS transition, in the presence of spin anisotropy. Given the physical meaning of the anisotropy term $\Delta$, we can identify the symmetry gapped phase to the easy-axis anisotropy ($\Delta<0$) and the first-order transition to the easy-plane anisotropy ($\Delta>0$). A schematic phase diagram is presented in \figref{fig:easy_plane} for the lattice model \eqnref{eq:EPJQ}. The symmetric gapped phase may as well be interpreted as the Ising ordered phase of $\langle S^z\rangle\neq 0$, since the condensation of $\phi_1$ field corresponds to the ordering of $\langle S^z\rangle$.
{The scenario that the AFM-VBS transition becomes first-order as the easy-plane anisotropy is turned on is consistent with the recent QMC study \refcite{Desai2020First-order}.}

\section{Summary}

In this work, we studied the \ntqed with self-duality. The \ntqed has $\SO(4)$ symmetry in the IR, if imposing the self-duality symmetry, it can be enhanced to $\Oo(4)$. The singlet mass is invariant under $\SO(4)$ but self-duality odd and the triplet mass is transformed by $\SO(4)$ but self-duality even. Requiring the $\Oo(4)$ symmetry, the theory cannot have explicit mass terms, which enables us to treat the mass terms as fluctuation scalar fields and to investigate the continuous-to-first-order transition driven by the mass fluctuations. The multi-critical points (lines) separating the continuous and first-order transitions can be described by the QED-GN theory.

We further analyzed the stability of the theory under the four-fermion interactions. In particular, we focus on the spatial current-current interaction of fermions in the field theory, which corresponds to a class of dimer-dimer interaction (the stagger-$Q$ term) in the lattice spin model\citen{zhao_multicritical_2020,zhao_tunable_2020}. This operator has been shown to drive the continuous DQCP to a first-order transition in recent numerical works. Our analysis indicates that such dimer interaction can be relevant at the $\Oo(4)$ DQCP and adjacent multi-critical lines, which generally destabilize the continuous DQCP to first-order transitions (or intermediate gapped phases). Our finding provides a theoretical understanding of the numerically observed first-order transition driven by the dimer-dimer interaction. Our analysis also suggests a possibility to have $\Z_2$ spin liquid in this model\citen{yang2021quantum}.

We provide systematically large-$N$ renormalization group calculation of the general $N_f=2$ QED$_3$ with Gross-Neveu term in \appref{appendix:largeN}. Thanks to viewing the Feynman diagrams as string diagrams of symmetry group representations\cite{cvitanovic2008group}, the complicated diagram at $\mathcal{O}(1/N_f)$ can be expressed by a few group parameters. Scaling dimensions of generic fermion/boson bilinear terms and four-fermion perturbations are presented. We expect these general results will find broader applications in other exotic quantum critical systems.

\begin{acknowledgments}
We acknowledge the helpful discussion with John McGreevy and Anders Sandvik. D.C.L. and Y.Z.Y. are supported by a startup fund at UCSD. C.X. is supported by NSF Grant No. DMR-1920434, and the Simons Foundation.
\end{acknowledgments}

\bibliography{bibbb}
\onecolumngrid
\newpage
\appendix
\section{Large-$N$ renormalization group}\label{appendix:largeN}
The theory considered in the main text is the QED$_3$ with level-$k$ Chern-Simons term and Yukawa coupling between the fermion bilinear terms and the scalar fields,
\begin{equation}\label{eq:general_model}
\begin{split}
&\mathcal{L}=\bar{\psi} (\mathds{1}_{N_f}\otimes \gamma^{\mu})(\partial_\mu -\mathsf{i}a_\mu ) \psi + \phi_a \bar{\psi} (M^a \otimes \mathds{1}_2) \psi\\
&+ \frac{1}{2 g^2} \phi_a (r_a-\partial^2)\phi_a +\frac{\lambda}{4}(\phi_a \phi_a)^2 \\
&+ \frac{\mathsf{i}k}{4\pi} \epsilon^{\mu \nu \lambda} a_\mu \partial_\nu a_\lambda+\frac{1}{4e^2}f_{\mu\nu}f^{\mu\nu}\\
\end{split}
\end{equation}
where $\psi, \bar{\psi}$ represents $N_f$ flavors of 2-component Dirac fermion fields, $\mathds{1}_N,M^a$ act on the $N_f$-dimensional flavor space while $\mathds{1}_2,\gamma^\mu,\Gamma^{(m),I}$ act on the 2-dimensional spinor space. $\Gamma^{(m),\mu_1,...,\mu_m}$ is defined as $\gamma^{[\mu_1}...\gamma^{\mu_m]}$ (antisymmetrize the indices) and any product of $\gamma$ matrices can be reduced to this form. Since the spacetime dimension is 3, $\Gamma^{(i)}$ and $\Gamma^{(3-i)}$ are related by the Levi-Civita tensor. $\phi_a$ with $a=1,...,N_b$ represent the scalar fields which are coupled to the fermion bilinears via a Yukawa type interaction. The last term in the Lagrangian is the Chern-Simons term with level $k$.

The bare propagators and vertices can be read off from the Lagrangian Eq.~\eqref{eq:general_model},
\begin{fmffile}{bare_propagator}
\begin{equation}
\begin{split}
    &\begin{gathered}
        \begin{fmfgraph*}(50,10)
        \fmfleft{i1}
        \fmfright{o1}
        \fmf{fermion}{i1,o1}
        \end{fmfgraph*}
    \end{gathered} =-\mathsf{i}\frac{p_\mu(\mathds{1}_{N_f}\otimes \gamma^\mu)}{p^2}, \quad \quad
    \begin{gathered}
        \begin{fmfgraph*}(50,10)
        \fmfleft{i1}
        \fmfright{o1}
        \fmf{dashes}{i1,o1}
        \fmflabel{$a$}{i1}
        \fmflabel{$b$}{o1}
        \end{fmfgraph*}
    \end{gathered} \quad =D^{(0)}_{ab}(q) = \frac{g^2}{q^2}\delta_{a,b} \\
    &\begin{gathered}
        \begin{fmfgraph*}(50,10)
        \fmfleft{i1}
        \fmfright{o1}
        \fmf{photon}{i1,o1}
        \fmflabel{$\mu$}{i1}
        \fmflabel{$\nu$}{o1}
        \end{fmfgraph*}
    \end{gathered} \quad =\Pi^{(0)}_{\mu\nu}(q)=\frac{e^2}{q^2}\left(\frac{q^2\delta_{\mu\nu}-q_\mu q_\nu+\frac{k}{2\pi} e^2\epsilon_{\mu\nu\rho}q^\rho}{q^2+(\frac{k}{2\pi})^2 e^4}+\xi \frac{q_\mu q_\nu}{q^2}\right)
\end{split}
\end{equation}
\end{fmffile}
where $\xi$ is the gauge parameter. The vertices are,
\begin{fmffile}{vertices}
\begin{equation}
    \begin{gathered}
        \begin{fmfgraph*}(50,40)
        \fmfleft{i1}
        \fmfright{o1,o2}
        \fmf{photon}{i1,v1}
        \fmf{fermion}{o2,v1,o1}
        \fmflabel{$\mu$}{i1}
        \phov{v1}
        \end{fmfgraph*}
    \end{gathered} = \ii \mathds{1}_{N_f} \otimes \gamma^\mu, \quad \quad
    \begin{gathered}
        \begin{fmfgraph*}(50,40)
        \fmfleft{i1}
        \fmfright{o1,o2}
        \fmf{dashes}{i1,v1}
        \fmf{fermion}{o2,v1,o1}
        \fmflabel{$a$}{i1}
        \bosv{v1}
        \end{fmfgraph*}
    \end{gathered} \quad =M^a \otimes \mathds{1}_2
\end{equation}
\end{fmffile}

The bare gauge and critical boson propagator will receive corrections, in the large-$N$ limit, the corrections are dominated by fermion loops, for the gauge propagator,
\begin{fmffile}{fermion_loop_gauge}
\begin{equation}
\begin{split}
    \begin{gathered}
        \begin{fmfgraph*}(50,40)
        \fmfleft{i}
        \fmfright{o}
        \fmf{photon}{i,v1}
        \fmf{photon}{v2,o}
        \fmf{fermion,left,tension=.3,label=$k+q$}{v1,v2}
        \fmf{fermion,left,tension=.3,label=$k$}{v2,v1}
        \fmflabel{$\mu$}{i} \fmflabel{$\nu$}{o}
        \phov{v1,v2}
        \end{fmfgraph*}
    \end{gathered} \quad &=(-1)[\ii \mathds{1}_{N_f} \otimes \gamma^\mu][-\mathsf{i}\frac{k_\rho(\mathds{1}_{N_f}\otimes \gamma^\rho)}{k^2}][\ii \mathds{1}_{N_f} \otimes \gamma^\nu][-\mathsf{i}\frac{(k+q)_\sigma(\mathds{1}_{N_f}\otimes \gamma^\sigma)}{(k+q)^2}]\\
    &= (-1) \tr [\mathds{1}_{N_f}\otimes \gamma^\mu\gamma^\rho\gamma^\nu\gamma^\sigma] \int \frac{d^3 k}{(2\pi)^3} \frac{k_\rho (k+q)_\sigma}{k^2 (k+q)^2}\\
    &= -\frac{N_f \abs{q}}{16}(\delta_{\mu\nu}-\frac{q_\mu q_\nu}{q^2})
\end{split}
\end{equation}
\end{fmffile}
where $N_f$ comes from trace over the identity matrix $\mathds{1}_{N_f}$. Similar for the critical boson propagator,
\begin{fmffile}{fermion_loop_boson}
\begin{equation}
\begin{split}
    \begin{gathered}
        \begin{fmfgraph*}(50,40)
        \fmfleft{i}
        \fmfright{o}
        \fmf{dashes}{i,v1}
        \fmf{dashes}{v2,o}
        \fmf{fermion,left,tension=.3,label=$k+q$}{v1,v2}
        \fmf{fermion,left,tension=.3,label=$k$}{v2,v1}
        \fmflabel{$a$}{i} \fmflabel{$b$}{o}
        \bosv{v1,v2}
        \end{fmfgraph*}
    \end{gathered} \quad &=(-1)[M^a \otimes \mathds{1}_2][-\mathsf{i}\frac{k_\rho(\mathds{1}_{N_f}\otimes \gamma^\rho)}{k^2}][M^b \otimes \mathds{1}_2][-\mathsf{i}\frac{(k+q)_\sigma(\mathds{1}_{N_f}\otimes \gamma^\sigma)}{(k+q)^2}]\\
    &= \tr [M^a M^b\otimes \gamma^\rho\gamma^\sigma] \int \frac{d^3 k}{(2\pi)^3} \frac{k_\rho (k+q)_\sigma}{k^2 (k+q)^2}\\
    &= -\tr[M^a M^b] \frac{\abs{q}}{8} \equiv -\mathbb{M} \frac{\abs{q}}{8} \delta_{ab}
\end{split}
\end{equation}
\end{fmffile}
where in the last step we define $\tr[M^a M^b]=\mathbb{M}\delta_{ab}$, this is true when $M^a$ is irreducible representation. The corrected propagator can be found by using Dyson's equation,
\begin{equation*}
    \Pi(q) = \{[\Pi^{(0)}(q)]^{-1}-\Sigma^{(0)}(q)\}^{-1}\\
\end{equation*}
Note that in the large-$N$ limit, this model flows to an interacting conformal field theory in the infrared limit, where the momentum scale $q$ is much smaller than the coupling constants $e,g$, therefore the leading order of the dressed gauge and critical boson propagators are,

\begin{fmffile}{dressed_propagator}
\begin{align}
\begin{split}
    \begin{gathered}
        \begin{fmfgraph*}(50,30)
        \fmfleft{i1} \fmfright{o1} \fmf{dbl_wiggly}{i1,o1} \fmflabel{$\mu$}{i1} \fmflabel{$\nu$}{o1}
        \end{fmfgraph*}
    \end{gathered}\quad &= \quad
    \begin{gathered}
        \begin{fmfgraph*}(50,30)
        \fmfleft{i1} \fmfright{o1} \fmf{wiggly}{i1,o1} \fmflabel{$\mu$}{i1} \fmflabel{$\nu$}{o1}
        \end{fmfgraph*}
    \end{gathered}\quad + \quad
    \begin{gathered}
        \begin{fmfgraph*}(50,30)
        \fmfleft{i} \fmfright{o} \fmf{photon}{i,v1} \fmf{photon}{v2,o} \fmf{fermion,left,tension=.3}{v1,v2,v1}
        \fmflabel{$\mu$}{i} \fmflabel{$\nu$}{o}
        \phov{v1,v2}
        \end{fmfgraph*}
    \end{gathered}\quad + \quad
    \begin{gathered}
        \begin{fmfgraph*}(70,30)
        \fmfleft{i} \fmfright{o}
        \fmf{photon}{i,v1} \fmf{photon}{v2,v3} \fmf{photon}{v4,o} \fmf{fermion,left,tension=.3}{v1,v2,v1} \fmf{fermion,left,tension=.3}{v3,v4,v3}
        \fmflabel{$\mu$}{i} \fmflabel{$\nu$}{o}
        \phov{v1,v2,v3,v4}
        \end{fmfgraph*}
    \end{gathered}\\
    \Pi_{\mu\nu}(q) & \simeq \frac{\mathbb{A}}{{N_f} \abs{q}}\left(\delta_{\mu\nu}-\xi\frac{q_\mu q_\nu}{q^2}\right)+\frac{\mathbb{B}}{{N_f}}\frac{\epsilon^{\mu\nu\rho}q_\rho}{q^2} +\mathcal{O}(\abs{q}/e^2)
\end{split}\\
\begin{split}
    \begin{gathered}
        \begin{fmfgraph*}(50,30)
        \fmfleft{i1} \fmfright{o1} \fmf{dbl_dashes}{i1,o1} \fmflabel{$a$}{i1} \fmflabel{$b$}{o1}
        \end{fmfgraph*}
    \end{gathered}\quad &= \quad
    \begin{gathered}
        \begin{fmfgraph*}(50,30)
        \fmfleft{i1} \fmfright{o1} \fmf{dashes}{i1,o1} \fmflabel{$a$}{i1} \fmflabel{$b$}{o1}
        \end{fmfgraph*}
    \end{gathered}\quad + \quad
    \begin{gathered}
        \begin{fmfgraph*}(50,30)
        \fmfleft{i} \fmfright{o} \fmf{dashes}{i,v1} \fmf{dashes}{v2,o} \fmf{fermion,left,tension=.3}{v1,v2,v1}
        \fmflabel{$a$}{i} \fmflabel{$b$}{o}
        \bosv{v1,v2}
        \end{fmfgraph*}
    \end{gathered}\quad + \quad
    \begin{gathered}
        \begin{fmfgraph*}(70,30)
        \fmfleft{i} \fmfright{o}
        \fmf{dashes}{i,v1} \fmf{dashes}{v2,v3} \fmf{dashes}{v4,o} \fmf{fermion,left,tension=.3}{v1,v2,v1} \fmf{fermion,left,tension=.3}{v3,v4,v3}
        \fmflabel{$a$}{i} \fmflabel{$b$}{o}
        \bosv{v1,v2,v3,v4}
        \end{fmfgraph*}
    \end{gathered}\\
    D_{ab}(q)& \simeq \frac{8}{\mathbb{M}\abs{q}}\delta_{ab} +\mathcal{O}(\abs{q}/g^2) \equiv D(q) \delta_{ab}
\end{split}
\end{align}
\end{fmffile}
where $\mathbb{A}=\left(16^{-1}+16\kappa^2\right)^{-1}, \mathbb{B}=\left((256\kappa)^{-1}+\kappa\right)^{-1}$, and $\kappa =k/(2\pi {N_f})$, a simple check is when $k=0$, $\mathbb{A}=16,\mathbb{B}=0$ match the coefficients in the large-$N$ analysis of QED$_3$ theory. Note that $\kappa$ is not inverse proportional to the 't Hooft coupling and can be any real number, the large-$N$ limit is to take ${N_f},k$ to $\infty$ while keeping $\kappa$ fixed. We also keep the gauge parameter $\xi$ in the calculation and check that the final result does not depend on $\xi$.

\subsection{Basic diagrams for $1/N$ corrections: self-energy}
We extract the logarithmic divergences from the diagrams and using $k,\Lambda$ to denote the external momentum and UV cutoff respectively, the self-energy corrections are,
\begin{fmffile}{self_energy_gauge}
\begin{align}
\begin{split}
    \begin{gathered}
        \begin{fmfgraph*}(50,40)
        \fmfleft{i} \fmfright{o} \fmf{fermion}{i,o} \fmf{dbl_wiggly,right,tension=0.3}{o,i}
        \phov{i,o}
        \end{fmfgraph*}
    \end{gathered} \quad &=\int\frac{d^3q}{(2\pi)^3} [\photonprop{\mu}{\nu}][\photonvert{\mu}][\fermionprop{k+q}{\si}][\photonvert{\nu}] \\
    &= (\photonvert{\mu}) k_\mu \frac{\mathbb{A}(1-3\xi)}{6\pi^2{N_f}}\ln(k/\Lambda) +reg.
\end{split}\\
\begin{split}
    \begin{gathered}
        \begin{fmfgraph*}(50,40)
        \fmfleft{i} \fmfright{o} \fmf{fermion}{i,o} \fmf{dbl_dashes,right,tension=0.3}{o,i}
        \bosv{i,o}
        \end{fmfgraph*}
    \end{gathered} \quad &=\int\frac{d^3q}{(2\pi)^3} [\bosonprop{a}{b}][\bosonvert{a}][\fermionprop{k+q}{\si}][\bosonvert{b}] \\
    &= (\photonvert{\mu}) k_\mu \frac{8 C_M}{6\pi^2\mathbb{M}}\ln(k/\Lambda) +reg.
\end{split}
\end{align}
\end{fmffile}
where we define $M^a M^a=C_M \mathds{1}_{N_f}$ in analogy of the Casimir.

\subsection{Basic diagrams for $1/N$ corrections: vertex corrections}
The four-fermion interactions in general can be added to the Lagrangian perturbatively, and assuming the small four-fermion perturbations won't drive the system to other fixed points. The general form for such interactions is,
\begin{equation*}
    K_{(\al,(m_1),I),(\be,(m_2),J)} \bpsi(V^\al\otimes \Gamma^{(m_1),I})\psi \bpsi(V^\be\otimes \Gamma^{(m_2),J})\psi
\end{equation*}
For simplicity and physical relevance, we will consider a subset of the four-fermion interactions with the form,
\begin{equation}\label{eq:inter_diag}
    \mathcal{L} \supset \mathcal{L}_\mathrm{int} = u_{\al,(m),I}(\bpsi (V^\al\otimes \Gamma^{(m),I})\psi)^2
\end{equation}

We introduce the diagrams for the interaction vertices as,
\begin{fmffile}{interaction_vertices}
\begin{equation}\label{eq:interaction_vertices}
    \begin{gathered}
        \begin{fmfgraph*}(60,30)
        \fmfleft{i1,i2}
        \fmfright{o1,o2}
        \fmf{fermion}{i2,vp2,o2}
        \fmf{phantom}{i1,vp1,o1}
        \fmffreeze
        \fmf{dots}{vp1,vp2}
        \intv{vp2}
        \end{fmfgraph*}
    \end{gathered} = u_{\al,(m),I} (V^\al\otimes \Gamma^{(m),I}), \quad \quad
    \begin{gathered}
        \begin{fmfgraph*}(60,30)
        \fmfleft{i1,i2}
        \fmfright{o1,o2}
        \fmf{fermion}{i2,vp2,o2}
        \fmf{fermion}{i1,vp1,o1}
        \fmffreeze
        \fmf{dots}{vp1,vp2}
        \intv{vp2,vp1}
        \end{fmfgraph*}
    \end{gathered} = K_{(\al,(m_1),I),(\be,(m_2),J)} \begin{bmatrix} (V^\al\otimes \Gamma^{(m_1),I}) \\ \otimes \\(V^\be\otimes \Gamma^{(m_2),J}) \end{bmatrix}
\end{equation}
\end{fmffile}

The vertex corrections are
\begin{fmffile}{vertex_correction}
\begin{align}
    \begin{gathered}
        \begin{fmfgraph*}(50,40)
        \fmfleft{i1} \fmfright{o1} \fmf{fermion}{i1,v,o1} \fmf{dbl_wiggly,right,tension=0.3}{o1,i1}
        \phov{i1,o1} \intv{v}
        \end{fmfgraph*}
    \end{gathered} \quad &= u_{\al,(m),I} (V^\al\otimes \Ga^{(m),I}) \frac{\mathbb{A}(-3-2C_{\gamma,(m),I} +3\xi)}{6\pi^2 {N_f}}\reg\\
    \begin{gathered}
        \begin{fmfgraph*}(50,40)
        \fmfleft{i1} \fmfright{o1} \fmf{fermion}{i1,v,o1} \fmf{dbl_dashes,right,tension=0.3}{o1,i1}
        \bosv{i1,o1} \intv{v}
        \end{fmfgraph*}
    \end{gathered} \quad &= u_{\al,(m),I} (V^\al\otimes \Ga^{(m),I}) \frac{8C_{\gamma,(m),I}C_{M,\al}}{6\pi^2\mathbb{M}}\reg
\end{align}
\end{fmffile}
where we define $C_{M,\al},C_{\gamma,(m),I}$ as $M^a V^\al M^a \equiv C_{M,\al} V^\al$ and $\gamma^\mu \Ga^{(m),I}\gamma^\mu \equiv C_{\gamma,(m),I} \Ga^{(m),I}$, repeated indices $a,\mu$ mean summation.\\
Note that there are also two-loop diagrams for the $1/N$ corrections, we begin with the calculation of the mass bubbles,
\begin{fmffile}{mass_bubbles}
\begin{align}
    &\begin{gathered}
        \begin{fmfgraph*}(50,40)
        \fmfstraight
        \fmfleft{i1,i2,i3} \fmfright{o1,o2,o3}
        \fmftop{t1}
        \fmf{fermion}{i1,o1,t1,i1}
        \intv{t1}  \fmfv{d.sh=circle,d.f=hatched,d.si=5}{i1,o1}
        \end{fmfgraph*}
    \end{gathered} \quad \begin{split} \Pi_{L,R}^\mathrm{Anticlockwise} = -\ii u_{\al,(m),I} &\tr[(V_L)(\mathds{1}_{N_f}\otimes \ga^a)(V_R)(V^\al \otimes(\ga^b \Ga^{(m),I} \ga^c))] \\
    &\frac{1}{128q^3}(q_a q_b q_c-q^2 q_c \delta_{a,b}-q^2 q_b \delta_{a,c}+q^2 q_a \delta_{b,c}), \end{split}\\
    &\begin{gathered}
        \begin{fmfgraph*}(50,40)
        \fmfstraight
        \fmfleft{i1,i2,i3} \fmfright{o1,o2,o3}
        \fmftop{t1}
        \fmf{fermion}{i1,t1,o1,i1}
        \intv{t1}  \fmfv{d.sh=circle,d.f=hatched,d.si=5}{i1,o1}
        \end{fmfgraph*}
    \end{gathered} \quad \begin{split} \Pi_{L,R}^\mathrm{Clockwise} = \ii u_{\al,(m),I} &\tr[(V_L)(V^\al \otimes(\ga^b \Ga^{(m),I} \ga^c)(V_R)(\mathds{1}_{N_f}\otimes \ga^a))] \\
    &\frac{1}{128q^3}(q_a q_b q_c-q^2 q_c \delta_{a,b}-q^2 q_b \delta_{a,c}+q^2 q_a \delta_{b,c}), \end{split}
\end{align}
\end{fmffile}
where $V_L,V_R$ stands for the vertex insertion, they could be gauge-gauge, boson-boson or gauge-boson. The formula is complicated in general. For the $3$-dimensional theory, the $\gamma$-matrices are simply the Pauli matrices and $m$ in $\Gamma^{(m),I}$ is up to $3$. Besides, $\Ga^{(3),\{i_1,i_2,i_3\}}=\ii \ep^{i_1,i_2,i_3}\Ga^{(0)}$, $\Ga^{(2),\{i_1,i_2\}}=\ii \ep^{i_1,i_2}_{\quad\;\; l}\Ga^{(1),l}$.

\paragraph{The gauge-gauge insertion:} Only $\Ga^{(0)},\Ga^{(3)}$ will have non-zero contribution, as their relation is discussed previously, we can calculate $\Ga^{(0)}$ and derive the result for $\Ga^{(3)}$. The mass bubble result for $\Ga^{(0)}$ is,
\begin{fmffile}{gauge-gauge-mass-bubl}
\begin{align}
    \begin{gathered}
        \begin{fmfgraph*}(30,40)
        \fmfstraight
        \fmfleft{i1,i2} \fmfright{o1,o2}
        \fmfpoly{shaded}{i1,o1,t1}
        \intv{t1} \phov{i1,o1}
        \end{fmfgraph*}
    \end{gathered} \quad = u_{\al,(0)} \tr(V) \frac{q_l \epsilon^{lij}}{4 q}
\end{align}
\end{fmffile}
\begin{fmffile}{gauge-gauge-2-loop}

The two-loop diagrams give similar results for $\Ga^{(0)},\Ga^{(3)}$,
\begin{align}
    \begin{gathered}
        \begin{fmfgraph*}(30,30)
        \fmfstraight
        \fmfleft{i0,i,i1,i2} \fmfright{o0,o,o1,o2}
        \fmfpoly{shaded}{i1,o1,t1}
        \fmf{dbl_wiggly}{i0,i1}
        \fmf{dbl_wiggly}{o0,o1}
        \fmf{fermion}{i0,o0}
        \intv{t1} \phov{i1,o1,i0,o0}
        \end{fmfgraph*}
    \end{gathered} \quad =  \begin{dcases}u_{\al,(0)}  (\mathds{1}_{N_f} \otimes \Ga^{(0)}) \frac{\tr(V)}{{N_f}}\frac{(\mathbb{A}^2-\mathbb{B}^2)}{4 \pi^2 {N_f}}\reg & \text{for } m=0\\
    u_{\al,(3),\{i_1,i_2,i_3\}}  (\mathds{1}_{N_f} \otimes \Ga^{(3),\{i_1,i_2,i_3\}}) \frac{\tr(V)}{{N_f}}\frac{(\mathbb{A}^2-\mathbb{B}^2)}{4 \pi^2 {N_f}}\reg & \text{for } m=3
    \end{dcases}
\end{align}
\end{fmffile}

\paragraph{The boson-boson insertion:} The non-zero contributions will occur only if $\tr(M^i M^j V^\al) = - \tr(M^i V^\al M^j)$, this requires non-trivial choices of the $M^a,V^\al$. If so, the two-loop contributions are,
\begin{fmffile}{boson-bonson-mass-bubl}
\begin{align}
    &\text{for }\tr(M^i M^j V^\al) = - \tr(M^i V^\al M^j) \nonumber\\
    &\begin{gathered}
        \begin{fmfgraph*}(30,40)
        \fmfstraight
        \fmfleft{i1,i2} \fmfright{o1,o2}
        \fmfpoly{shaded}{i1,o1,t1}
        \intv{t1} \bosv{i1,o1}
        \end{fmfgraph*}
    \end{gathered} \quad =
    \begin{dcases} -\ii u_{\al,(1),i_1} \tr(M^i V^\al M^j) \frac{q_{i_1}}{4q} & \text{for }m=1\\
    -\ii u_{\al,(2),\{i_1,i_2\}} \tr(M^i V^\al M^j) \frac{\ii q_{l}\epsilon^{l,i_1,i_2}}{2q} & \text{for }m=2
    \end{dcases}
\end{align}
\end{fmffile}
\begin{fmffile}{boson-boson-2-loop}
\begin{align}
    \begin{gathered}
        \begin{fmfgraph*}(30,30)
        \fmfstraight
        \fmfleft{i0,i,i1,i2} \fmfright{o0,o,o1,o2}
        \fmfpoly{shaded}{i1,o1,t1}
        \fmf{dbl_dashes}{i0,i1}
        \fmf{dbl_dashes}{o0,o1}
        \fmf{fermion}{i0,o0}
        \intv{t1} \bosv{i1,o1,i0,o0}
        \end{fmfgraph*}
    \end{gathered} \quad =  \begin{dcases} u_{\al,(1),i_1} (M^i M^j \otimes \Ga^{(1),i_1}) \frac{-8\tr(M^i V^\al M^j)}{3\pi^2 \mathbb{M}^2}\reg & \text{for }m=1\\
    u_{\al,(2),\{i_1,i_2\}} (M^i M^j \otimes \Ga^{(2),\{i_1,i_2\}}) \frac{-8\tr(M^i V^\al M^j)}{3\pi^2 \mathbb{M}^2}\reg & \text{for }m=2
    \end{dcases}
\end{align}
\end{fmffile}
For example, the boson 2-loop will contribute when $M^a=\{\dsi_2,\sigma^1,\sigma^2\}$ and $V^\al=\{\dsi_2,\sigma^1,\sigma^2,\sigma^3\}$ and it will only contribute to the current-current interaction.

\paragraph{Mixed gauge-boson insertion} The mixed gauge-boson insertion will vanish for all the choices of $V^\al$ and $\Ga^{(m),I}$, part of the reason is because $\tr(M^i V^\al) = \tr(V^\al M^i)$ and it will never have a minus sign.

\subsection{Basic diagrams for $1/N$ corrections: ladder corrections}
The four-fermion interaction vertices as depicted in Eq.~\eqref{eq:interaction_vertices} will receive $\mathcal{O}(1/N)$ correction from gauge and boson propagators as well,
\begin{fmffile}{ladder_correction_gauge}
\begin{align}\label{eq:ladder_g}
    &\begin{gathered}
        \begin{fmfgraph*}(80,30)
        \fmfleft{i1,i2}
        \fmfright{o1,o2}
        \fmf{fermion}{i2,vp2,vq2,o2}
        \fmf{fermion}{i1,vp1,vq1,o1}
        \fmffreeze
        \fmf{dbl_wiggly}{vq1,vq2}
        \fmf{dots}{vp1,vp2}
        \intv{vp2,vp1} \phov{vq1,vq2}
        \end{fmfgraph*}
    \end{gathered} +
    \begin{gathered}
        \begin{fmfgraph*}(80,30)
        \fmfleft{i1,i2}
        \fmfright{o1,o2}
        \fmf{fermion}{i2,vp2,vq2,o2}
        \fmf{fermion}{i1,vp1,vq1,o1}
        \fmffreeze
        \fmf{dbl_wiggly}{vp1,vq2}
        \fmf{dots}{vp2,vq1}
        \intv{vp2,vq1} \phov{vp1,vq2}
        \end{fmfgraph*}
    \end{gathered} \nonumber \\
    &= K_{(\al,(m_1),I),(\be,(m_2),J)} \begin{bmatrix} V^\al \\ \otimes \\V^\be \end{bmatrix} \otimes \left(\begin{bmatrix} \Ga^{(m_1),I}\ga^\mu \\ \otimes \\\Ga^{(m_2),J}\ga^\mu \end{bmatrix} + \begin{bmatrix} \Ga^{(m_1),I}\ga^\mu \\ \otimes \\\ga^\mu\Ga^{(m_2),J} \end{bmatrix}\right) \frac{-2\mathbb{A}}{6\pi^2 {N_f}}\reg\\
    &\Rightarrow u_{(\al,(m),I)} \begin{bmatrix} V^\al \\ \otimes \\V^\al \end{bmatrix} \otimes \left(\begin{bmatrix} \Ga^{(m),I}\ga^\mu \\ \otimes \\\Ga^{(m),I}\ga^\mu \end{bmatrix} + \begin{bmatrix} \Ga^{(m),I}\ga^\mu \\ \otimes \\\ga^\mu\Ga^{(m),I} \end{bmatrix}\right) \frac{-2\mathbb{A}}{6\pi^2 {N_f}}\reg
\end{align}
\end{fmffile}
where the last equation is the correction for the simplified four-fermion interaction as in Eq.~\ref{eq:inter_diag}.

\begin{fmffile}{ladder_correction_boson}
\begin{align}\label{eq:ladder_b}
    &\begin{gathered}
        \begin{fmfgraph*}(80,30)
        \fmfleft{i1,i2}
        \fmfright{o1,o2}
        \fmf{fermion}{i2,vp2,vq2,o2}
        \fmf{fermion}{i1,vp1,vq1,o1}
        \fmffreeze
        \fmf{dbl_dashes}{vq1,vq2}
        \fmf{dots}{vp1,vp2}
        \intv{vp2,vp1} \bosv{vq1,vq2}
        \end{fmfgraph*}
    \end{gathered} +
    \begin{gathered}
        \begin{fmfgraph*}(80,30)
        \fmfleft{i1,i2}
        \fmfright{o1,o2}
        \fmf{fermion}{i2,vp2,vq2,o2}
        \fmf{fermion}{i1,vp1,vq1,o1}
        \fmffreeze
        \fmf{dbl_dashes}{vp1,vq2}
        \fmf{dots}{vp2,vq1}
        \intv{vp2,vq1} \bosv{vp1,vq2}
        \end{fmfgraph*}
    \end{gathered} \nonumber \\
    &= K_{(\al,(m_1),I),(\be,(m_2),J)} \left(\begin{bmatrix} V^\al M^a \\ \otimes \\V^\be M^a \end{bmatrix} \otimes\begin{bmatrix} \Ga^{(m_1),I}\ga^\mu \\ \otimes \\\Ga^{(m_2),J}\ga^\mu \end{bmatrix} -
    \begin{bmatrix} V^\al M^a \\ \otimes \\M^a V^\be \end{bmatrix} \otimes\begin{bmatrix} \Ga^{(m_1),I}\ga^\mu \\ \otimes \\\ga^\mu\Ga^{(m_2),J} \end{bmatrix}\right) \frac{-8}{6\pi^2 \mathbb{M}}\reg\\
    &= u_{(\al,(m),I)} \left(\begin{bmatrix} V^\al M^a \\ \otimes \\V^\al M^a \end{bmatrix} \otimes\begin{bmatrix} \Ga^{(m),I}\ga^\mu \\ \otimes \\\Ga^{(m),I}\ga^\mu \end{bmatrix} -
    \begin{bmatrix} V^\al M^a \\ \otimes \\M^a V^\al \end{bmatrix} \otimes\begin{bmatrix} \Ga^{(m),I}\ga^\mu \\ \otimes \\\ga^\mu\Ga^{(m),I} \end{bmatrix}\right) \frac{-8}{6\pi^2 \mathbb{M}}\reg
\end{align}
\end{fmffile}
the repeated indices $\mu,a$ should be summed over.

\begin{table}[]
    \centering
    \begin{tabular}{c|c}
        Notation & Definition\\ \hline
        $\kappa$ & $k/(2\pi {N_f})$ \\ \hline
        $\mathbb{A}$ & $\mathbb{A}=\left(16^{-1}+16\kappa^2\right)^{-1}$ \\ \hline
        $\mathbb{B}$ & $\mathbb{B}=\left((256\kappa)^{-1}+\kappa\right)^{-1}$ \\ \hline
        $\mathbb{M}$ & $\tr[M^a M^b]=\mathbb{M}\delta_{ab}$ \\ \hline
        $C_M$ & $M^a M^a=C_M \mathds{1}_{N_f}$ \\ \hline
        $C_{M,\al}$ & $M^a V^\al M^a \equiv C_{M,\al} V^\al$\\ \hline
        $C_{\gamma,(m),I}$ & \makecell{$\gamma^\mu \Ga^{(m),I}\gamma^\mu \equiv C_{\gamma,(m),I} \Ga^{(m),I}$ \\ $C_{\gamma,(0),}=3,C_{\gamma,(1),\mu}=-1$} \\ \hline
        $f_{c}^{ab}$ &  $\si^a \si^b=f_c^{ab}\si^c,\ \frac{1}{2}\tr(\si^a \si^b \si^c)$ \\ \hline
        $\mF_\be^\al(\{M^a\}),\tmF_\be^\al(\{M^a\})$ & \makecell{$\mF_\be^\al(\{M^a\}) = \sum_{a\in \{M^a\}} f_{\be}^{\al a}f_{\be}^{\al a}$,\\  $\tmF_\be^\al(\{M^a\}) = \sum_{a\in \{M^a\}} f_{\be}^{\al a}f_{\be}^{a \al },\  \al,\be \in \{V^\al\}$}
    \end{tabular}
    \caption{The definition for the coefficients that are universal for chosen fermion-boson vertex and interaction matrix.}
    \label{tab:notations}
\end{table}

\subsection{Examples}
Above general calculations will be concrete with certain assumptions,
\begin{enumerate}
    \item Since $\Ga^{(i)}$ and $\Ga^{(3-i)}$ are related by Levi-Civita symbol in 3 dimension, we only need to consider $\Ga^{(0)}=\dsi_2$ and $\Ga^{(1),\mu}=\ga^\mu$.
    \item For the physical relevance, we consider the four-fermion interactions in the form of $u_{\al,m}(\bpsi(V^\al\otimes \dsi_2)\psi)^2$ and $u_{\al,\mu}(\bpsi(V^\al\otimes \ga^\mu)\psi)^2$ with $\mu=0,1,2$.
    \item We further assume $V^\al,M^a$ are represented by Pauli matrices. This kind of interaction vertices arise when doing fermionic parton construction of the spin models, i.e. the spin operators correspond to the fermion bilinears with Pauli matrices inserted in the middle.
    \item We also view $V^\al$s as the basis of certain vector space and form a set $\{V^\al\}$ as well as $M^a$s form a set $\{M^a\}$, such that $V^\al M^a \in \{V^\al\}$. For example, for $\{M^a\}=\{\dsi_2\}$ or $\{M^a\}=\{\dsi_2,\sigma^3\}$, $\{V^\al\}$ can be $\{V^\al\}=\{\dsi_2,\sigma^3\}$ or $\{V^\al\}=\{\dsi_2,\sigma^1,\sigma^2,\sigma^3\}$
\end{enumerate}
Since $V^\al,M^a$ can be represented by Pauli matrices as assumed, we can exploit the underlying algebraic structure of Pauli matrices. We further define the structure constants when multiplying the $V^\al,M^a$ as
\begin{equation}
    V^\al M^b=\sum_\be f_{\be}^{\al b}V^\be,\quad M^b V^\al = \sum_\be f_{\be}^{b\al}V^\be
\end{equation}
where $f_{\be}^{\al b},f_{\be}^{b\al}$ can be viewed as $\sigma^a\sigma^b=\sum_c f_{c}^{ab}\sigma^c$ with $a,b,c$ being restricted. The $f_{c}^{ab}$ for Pauli matrices are,
\begin{equation}\label{eq:structure_constant}
    f_{c}^{ab}=\ii \epsilon_c^{ab},\text{with } a,b,c=1,2,3,\quad f_{b}^{0a}=\delta_b^a,\ f_{b}^{a0}=\delta_b^a,\  f_{0}^{ab}=\delta^{ab}, \text{with } a,b=0,1,2,3.
\end{equation}
The structure constants are also calculated by,
\begin{equation}
    f_c^{ab}=\frac{1}{2}\tr(\si^a\si^b\si^c)
\end{equation}
The $\ga,\Ga$ matrices are also represented by Pauli matrices and therefore have this structure as well,
\begin{equation}
    \Ga^i \gamma^\mu=\sum_j f_{j}^{i\mu}\Ga^j,\quad  \gamma^\mu \Ga^i=\sum_j f_{j}^{\mu i}\Ga^j.
\end{equation}

We arrange the coupling constants in a vector as $u_{\al,i} = (u_{\al,m},u_{\al,0},u_{\al,1},u_{\al,2})$, where the first term is the mass-mass interaction and the last 3 terms are the current-current interactions in $\tau,x,y$ directions. The corresponding $\gamma$-matrices are $\Ga^i=\{\dsi_2,\ga^0,\ga^1,\ga^2\}=\{\dsi_2,\si^3,\si^1,\si^2\}$.

With the structure constants, the ladder corrections Eq.~\ref{eq:ladder_g} can be simplified as,
\begin{align}
    &u_{(\al,(m),I)} \begin{bmatrix} V^\al \\ \otimes \\V^\al \end{bmatrix} \otimes \left(\begin{bmatrix} \Ga^{(m),I}\ga^\mu \\ \otimes \\\Ga^{(m),I}\ga^\mu \end{bmatrix} + \begin{bmatrix} \Ga^{(m),I}\ga^\mu \\ \otimes \\\ga^\mu\Ga^{(m),I} \end{bmatrix}\right) \frac{-2\mathbb{A}}{6\pi^2 {N_f}}\reg\\
    =&u_{(\al,i)} \begin{bmatrix} V^\al \\ \otimes \\V^\al \end{bmatrix} \otimes \begin{bmatrix} \Ga^{j}\\ \otimes \\\Ga^{j}\end{bmatrix}\sum_{\mu}\left(f_{j}^{i\mu}f_{j}^{i\mu}+f_{j}^{i\mu}f_{j}^{\mu i}\right)  \frac{-2\mathbb{A}}{6\pi^2 {N_f}}\reg\\
    =&u_{(\al,i)} \begin{bmatrix} V^\al \\ \otimes \\V^\al \end{bmatrix} \otimes \begin{bmatrix} \Ga^{j}\\ \otimes \\\Ga^{j}\end{bmatrix}\left(\dsi_{\dim \{V^\al\}}\otimes\left(\mF_j^i+\tmF_j^i\right)\right)  \frac{-2\mathbb{A}}{6\pi^2 {N_f}}\reg
\end{align}
where $\mF_j^i\equiv \sum_{\mu=\{3,1,2\}} f_{j}^{i \mu}f_{j}^{i \mu}, \tmF_j^i \equiv \sum_{\mu=\{3,1,2\}} f_{j}^{i \mu}f_{j}^{\mu i },\ i,j=0,3,1,2$ and Eq.~\ref{eq:ladder_b} can be simplified as,
\begin{align}
    &u_{(\al,(m),I)} \left(\begin{bmatrix} V^\al M^a \\ \otimes \\V^\al M^a \end{bmatrix} \otimes\begin{bmatrix} \Ga^{(m),I}\ga^\mu \\ \otimes \\\Ga^{(m),I}\ga^\mu \end{bmatrix} -
    \begin{bmatrix} V^\al M^a \\ \otimes \\M^a V^\al \end{bmatrix} \otimes\begin{bmatrix} \Ga^{(m),I}\ga^\mu \\ \otimes \\\ga^\mu\Ga^{(m),I} \end{bmatrix}\right) \frac{-8}{6\pi^2 \mathbb{M}}\reg\\
    =& u_{(\al,i)} \begin{bmatrix} V^{\be}\\ \otimes \\V^{\be}\end{bmatrix} \otimes\begin{bmatrix} \Ga^{j}\\ \otimes \\\Ga^{j}\end{bmatrix}\left(\left( \sum_a f_{\be}^{\al a}f_{\be}^{\al a}\right)\left( \sum_\mu f_{j}^{i\mu}f_{j}^{i\mu}\right)-\left( \sum_a f_{\be}^{\al a}f_{\be}^{a\al }\right)\left( \sum_\mu f_{j}^{i\mu}f_{j}^{\mu i}\right)\right) \frac{-8}{6\pi^2 \mathbb{M}}\reg\\
    =& u_{(\al,i)} \begin{bmatrix} V^{\be}\\ \otimes \\V^{\be}\end{bmatrix} \otimes\begin{bmatrix} \Ga^{j}\\ \otimes \\\Ga^{j}\end{bmatrix}\left(\mF_\be^\al(\{M^a\})\otimes \mF_j^i-\tmF_\be^\al(\{M^a\})\otimes\tmF_j^i\right) \frac{-8}{6\pi^2 \mathbb{M}}\reg
\end{align}
Both of the ladder contributions will depend on the structure constants with specific forms, and we define $\mF_\be^\al(\{M^a\}) \equiv \sum_{a\in \{M^a\}} f_{\be}^{\al a}f_{\be}^{\al a}, \tmF_\be^\al(\{M^a\}) \equiv \sum_{a\in \{M^a\}} f_{\be}^{\al a}f_{\be}^{a \al },\  \al,\be \in \{V^\al\}$ similar to the above definition for the $\Ga$-matrices.

The self-energy corrections and the vertex corrections will be diagonal matrices acting on the vector $u_{\al,i}$. The self-energy corrections are the same for every $u_{\al,i}$, while the vertex corrections depend on the $\al,i$. As listed in the Table.~\ref{tab:notations}, the coefficient $C_{\gamma,(0),}=3,C_{\gamma,(1),\mu}=-1$ are distinct for mass-mass and current-current. The structure constants in Eq.~\ref{eq:structure_constant} also have these distinctions, this suggests the RG equations are in block forms. 

\subsection{Renormalization group equation for four-fermion interactions}
The $1/N$ corrections for the four-fermion interaction vertices are,

\begin{fmffile}{four-fermion-interaction-rg1}
\begin{equation}
\begin{split}
    -2\times \Bigg\{
    \begin{gathered}
        \fmfstraight
        \begin{fmfgraph*}(80,30)
        \fmftop{i2,va2,vb2,vc2,o2}
        \fmfbottom{i1,va1,vb1,vc1,o1}
        \fmf{fermion}{i2,va2,o2}
        \fmf{fermion}{i1,va1,vb1,vc1,o1}
        \fmffreeze
        \fmf{dots}{va1,va2}
        \fmf{dbl_wiggly,right,tension=0.3}{vb1,vc1}
        \intv{va1,va2} \phov{vb1,vc1}
        \end{fmfgraph*}
    \end{gathered} +
    \begin{gathered}
        \fmfstraight
        \begin{fmfgraph*}(70,30)
        \fmftop{i2,va2,vb2,vc2,o2}
        \fmfbottom{i1,va1,vb1,vc1,o1}
        \fmf{fermion}{i2,va2,o2}
        \fmf{fermion}{i1,va1,vb1,vc1,o1}
        \fmffreeze
        \fmf{dots}{vb1,vb2}
        \fmf{dbl_wiggly,right,tension=0.3}{va1,vc1}
        \intv{vb1,vb2} \phov{va1,vc1}
        \end{fmfgraph*}
    \end{gathered} +
    \left(\begin{gathered}
        \begin{fmfgraph*}(70,30)
        \fmfleft{i1,i2}
        \fmfright{o1,o2}
        \fmf{fermion}{i2,vp2,vq2,o2}
        \fmf{fermion}{i1,vp1,vq1,o1}
        \fmffreeze
        \fmf{dbl_wiggly}{vq1,vq2}
        \fmf{dots}{vp1,vp2}
        \intv{vp2,vp1} \phov{vq1,vq2}
        \end{fmfgraph*}
    \end{gathered} +
    \begin{gathered}
        \begin{fmfgraph*}(70,30)
        \fmfleft{i1,i2}
        \fmfright{o1,o2}
        \fmf{fermion}{i2,vp2,vq2,o2}
        \fmf{fermion}{i1,vp1,vq1,o1}
        \fmffreeze
        \fmf{dbl_wiggly}{vp1,vq2}
        \fmf{dots}{vp2,vq1}
        \intv{vp2,vq1} \phov{vp1,vq2}
        \end{fmfgraph*}
    \end{gathered}\right) +\begin{gathered}
        \begin{fmfgraph*}(70,50)
        \fmfstraight
        \fmfleft{i0,i1,i2,i3}
        \fmfright{o0,o1,o2,o3}
        \fmf{fermion}{i3,c3,o3}
        \fmf{fermion}{i0,v0,w0,o0}
        \fmf{phantom}{i2,c2,o2}
        \fmffreeze
        \fmf{dots}{c3,c2}
        \fmfpoly{shaded}{v1,w1,c2}
        \fmf{dbl_wiggly,tension=3}{v0,v1}
        \fmf{dbl_wiggly,tension=3}{w0,w1}
        \intv{c2,c3} \phov{v0,v1,w0,w1}
        \end{fmfgraph*}
    \end{gathered} \\
    + \begin{gathered}
        \fmfstraight
        \begin{fmfgraph*}(80,30)
        \fmftop{i2,va2,vb2,vc2,o2}
        \fmfbottom{i1,va1,vb1,vc1,o1}
        \fmf{fermion}{i2,va2,o2}
        \fmf{fermion}{i1,va1,vb1,vc1,o1}
        \fmffreeze
        \fmf{dots}{va1,va2}
        \fmf{dbl_dashes,right,tension=0.3}{vb1,vc1}
        \intv{va1,va2} \bosv{vb1,vc1}
        \end{fmfgraph*}
    \end{gathered} +
    \begin{gathered}
        \fmfstraight
        \begin{fmfgraph*}(70,30)
        \fmftop{i2,va2,vb2,vc2,o2}
        \fmfbottom{i1,va1,vb1,vc1,o1}
        \fmf{fermion}{i2,va2,o2}
        \fmf{fermion}{i1,va1,vb1,vc1,o1}
        \fmffreeze
        \fmf{dots}{vb1,vb2}
        \fmf{dbl_dashes,right,tension=0.3}{va1,vc1}
        \intv{vb1,vb2} \bosv{va1,vc1}
        \end{fmfgraph*}
    \end{gathered} +
    \left(\begin{gathered}
        \begin{fmfgraph*}(70,30)
        \fmfleft{i1,i2}
        \fmfright{o1,o2}
        \fmf{fermion}{i2,vp2,vq2,o2}
        \fmf{fermion}{i1,vp1,vq1,o1}
        \fmffreeze
        \fmf{dbl_dashes}{vq1,vq2}
        \fmf{dots}{vp1,vp2}
        \intv{vp2,vp1} \bosv{vq1,vq2}
        \end{fmfgraph*}
    \end{gathered} +
    \begin{gathered}
        \begin{fmfgraph*}(70,30)
        \fmfleft{i1,i2}
        \fmfright{o1,o2}
        \fmf{fermion}{i2,vp2,vq2,o2}
        \fmf{fermion}{i1,vp1,vq1,o1}
        \fmffreeze
        \fmf{dbl_dashes}{vp1,vq2}
        \fmf{dots}{vp2,vq1}
        \intv{vp2,vq1} \bosv{vp1,vq2}
        \end{fmfgraph*}
    \end{gathered}\right) +\begin{gathered}
        \begin{fmfgraph*}(70,50)
        \fmfstraight
        \fmfleft{i0,i1,i2,i3}
        \fmfright{o0,o1,o2,o3}
        \fmf{fermion}{i3,c3,o3}
        \fmf{fermion}{i0,v0,w0,o0}
        \fmf{phantom}{i2,c2,o2}
        \fmffreeze
        \fmf{dots}{c3,c2}
        \fmfpoly{shaded}{v1,w1,c2}
        \fmf{dbl_dashes,tension=3}{v0,v1}
        \fmf{dbl_dashes,tension=3}{w0,w1}
        \intv{c2,c3} \bosv{v0,v1,w0,w1}
        \end{fmfgraph*}
    \end{gathered}\Bigg\}.
\end{split}
\end{equation}
\end{fmffile}

As discussed previously, for generic boson-fermion vertices, the ladder correction diagram of one interaction vertex will contribute to another interaction vertex, therefore, one need to include all the possible interaction vertices as the basis. For example, if $\{M^a\}=\{\dsi_2,\sigma^3\}$ and $\{V^\al\}=\{\dsi_2,\sigma^1\}$, then $\sigma^2,\sigma^3$ also need to be included in $\{V^\al\}$.

We will analyze the example in the main text Section.~\ref{sec:scaling_4fermion} in detail. Due to the reason provided in the previous paragraph, we choose the interaction vertex to be $\{V^\al\}=\{\dsi_2,\sigma^1,\sigma^2,\sigma^3\}\otimes \dsi_N$ and $\{M^a\}=\{\},\{\dsi_2\}\otimes \dsi_N,\{\si^3\}\otimes \dsi_N,\{\dsi_2,\si^3\}\otimes \dsi_N$. Combining with the $\Gamma$-matrices, the basis of the interaction vertices $u_{\al,i} = (u_{\al,m},u_{\al,0},u_{\al,1},u_{\al,2})$ is $4\times 4=16$ dimensional. The RG equation is organized as,
\begin{equation}
    \frac{d u_{\al,i}}{d \ell} =\left(-\dsi+\frac{64}{3\pi^2 (2N)}\mM_{(\al,i),(\beta,j)}\right)u_{\beta,j}
\end{equation}
where $\al,\be$ are the indices of the flavors, and $i,j$ are the indices of the $\Ga$-matrices, $i=0,1,2,3$ corresponds to $\{\dsi_2,\ga^0,\ga^1,\ga^2\} =\{\dsi_2,\si^3,\si^1,\si^2\} $ in 3d.

The matrix $\mM_{(\al,i),(\beta,j)}$ contains several parts,
\begin{equation}
    \mM_{(\al,i),(\beta,j)}=\mM^{sv}_{(\al,i),(\beta,j)}+\mM^{gL}_{(\al,i),(\beta,j)}+\mM^{bL}_{(\al,i),(\beta,j)}
\end{equation}
The self-energy and vertex corrections are in the diagonal,
\begin{align}
    \mM^{sv}_{(\al,0),(\beta,0)} &=\left(\frac{16\mathbb{A}}{6\pi^2{N_f}}     +\frac{-2\tr(V^\al)}{{N_f}}\frac{(\mathbb{A}^2-\mathbb{B}^2)}{4 \pi^2 {N_f}} + \frac{-16(C_M+3C_{M,\al})}{6\pi^2\mathbb{M}}\right)\dsi_{\al,\be} \\
    \mM^{sv}_{(\al,i),(\beta,i)} &=\left(0 + \frac{-16(C_M-C_{M,\al})}{6\pi^2\mathbb{M}}\right)\dsi_{\al,\be},\quad \text{with } i=1,2,3
\end{align}
The ladder correction from the gauge vertex contributes the off-diagonal part,
\begin{equation}
    \mM^{gL}_{(\al,0),(\beta,i)}=\mM^{gL}_{(\al,i),(\beta,0)}= \frac{8\mathbb{A}}{6\pi^2 {N_f}}\dsi_{\al,\be},\quad \text{with } i=1,2,3.
\end{equation}
The ladder corrections from the boson vertex are complicated, in the $(i,j)$ space, there are two parts,
\begin{align}
    \mM^{bL}_{(\al,0),(\beta,i)} &= \mM^{bL}_{(\al,i),(\beta,0)} =\frac{16}{6\pi^2 \mathbb{M}}(\mF_\be^\al(\{M^a\})-\tmF_\be^\al(\{M^a\})),\quad \text{with } i=1,2,3\\
    \mM^{bL}_{(\al,i),(\beta,j)} &= \frac{16}{6\pi^2 \mathbb{M}}(-\mF_\be^\al(\{M^a\})-\tmF_\be^\al(\{M^a\})),\quad \text{with } i,j=1,2,3,i\ne j
\end{align}
where $\mF_\be^\al(\{M^a\}) = \sum_{a\in \{M^a\}} f_{\be}^{\al a}f_{\be}^{\al a}, \tmF_\be^\al(\{M^a\}) = \sum_{a\in \{M^a\}} f_{\be}^{\al a}f_{\be}^{a \al },\  \al,\be \in \{V^\al\}$ is defined previously. This can be simplified if we take subset of $\{V^\al\}$ with proper $\{M^a\}$, and restrict the indices $\al,\be$ in the subset.

\noindent\textbf{The first quadrant, continuous $\Oo(4)$ DQCP}: There is no critical boson in the system, $\{M^a\}=\{\}$. There is no mixture in the flavor space of the eigen-channel. For $V^a=\dsi_{2N}$,
\begin{align}
    \frac{d u_{0,i}}{d \ell} =\left(-1+\frac{64}{3\pi^2 (2N)} \mM_{(0,i),(0,j)}\right)u_{0,j},\quad
    \mM=\frac{1}{256 \kappa ^2+1}\left(
\begin{array}{cccc}
 \frac{4 \left(512 \kappa ^2-1\right)}{256 \kappa ^2+1} & 1 & 1 & 1 \\
 1 & 0 & 0 & 0 \\
 1 & 0 & 0 & 0 \\
 1 & 0 & 0 & 0 \\
\end{array}
\right)
\end{align}
In our case, $2N=2,\ka=0$, the RG equation becomes,
\begin{equation}
    \frac{d u_{0,i}}{d \ell} =\left(
\begin{array}{cccc}
 -1-\frac{128}{3 \pi ^2} & \frac{32}{3 \pi ^2} & \frac{32}{3 \pi ^2} & \frac{32}{3 \pi ^2} \\
 \frac{32}{3 \pi ^2} & -1 & 0 & 0 \\
 \frac{32}{3 \pi ^2} & 0 & -1 & 0 \\
 \frac{32}{3 \pi ^2} & 0 & 0 & -1 \\
\end{array}
\right)u_{0,j}
\end{equation}
And the eigenvalues of this matrix are all negative, meaning the perturbation is irrelevant among all the channels.

For $V^\al=\sigma^\al\otimes \dsi_N,\al=1,2,3$, the RG equations are the same for different $\al$s,
\begin{equation}\label{eq:RGequation_O(4)_tri}
    \frac{d u_{\al,i}}{d \ell} =\left(-1+\frac{64}{3\pi^2 (2N)} \mM_{(\al,i),(\al,j)}\right)u_{\al,j},\quad
    \mM=\frac{1}{256 \kappa ^2+1}\left(
\begin{array}{cccc}
 2 & 1 & 1 & 1 \\
 1 & 0 & 0 & 0 \\
 1 & 0 & 0 & 0 \\
 1 & 0 & 0 & 0 \\
\end{array}
\right)
\end{equation}
In our case, there is one relevant channel, and plug that into the Eq.~\ref{eq:RGequation_O(4)_tri}, we get,
\begin{align}\label{eq:O(4)_scaling_v3}
    u_{\al,i}=g_{\al}(3,1,1,1)^T, \quad \frac{d g_{\al}}{d \ell} = 2.24 g_{\al}.
\end{align}
Follow the same procedure, we will present the RG equations and the relevant channel results for other cases.

\vbox{}
\noindent\textbf{The $r_2$ axis, continuous $\SO(5)$ DQCP}: The boson corresponding to the singlet mass is critical, $\{M^a\}=\{\dsi_{2N}\}$. There is also no mixture in the flavor space. For $V^\al=\dsi_{2N}$,
\begin{align}
    \frac{d u_{0,i}}{d \ell} =\left(-1+\frac{64}{3\pi^2 (2N)} \mM_{(0,i),(0,j)}\right)u_{0,j},\quad
    \mM=\left(
\begin{array}{cccc}
 \frac{4 \left(512 \kappa ^2-1\right)}{\left(256 \kappa ^2+1\right)^2}-\frac{1}{2} & \frac{1}{256 \kappa ^2+1} & \frac{1}{256 \kappa ^2+1} & \frac{1}{256 \kappa ^2+1} \\
 \frac{1}{256 \kappa ^2+1} & 0 & -\frac{1}{4} & -\frac{1}{4} \\
 \frac{1}{256 \kappa ^2+1} & -\frac{1}{4} & 0 & -\frac{1}{4} \\
 \frac{1}{256 \kappa ^2+1} & -\frac{1}{4} & -\frac{1}{4} & 0 \\
\end{array}
\right).
\end{align}
There is no relevant channel in this case.

Again, for $V^\al=\sigma^\al\otimes \dsi_N,\al=1,2,3$, the RG equations are the same for different $\al$s,
\begin{equation}
    \frac{d u_{\al,i}}{d \ell} =\left(-1+\frac{64}{3\pi^2 (2N)} \mM_{(\al,i),(\al,j)}\right)u_{\al,j},\quad
    \mM=\left(
\begin{array}{cccc}
\frac{2}{256 \kappa ^2+1}-\frac{1}{2} & \frac{1}{256 \kappa ^2+1} & \frac{1}{256 \kappa ^2+1} & \frac{1}{256 \kappa ^2+1} \\
 \frac{1}{256 \kappa ^2+1} & 0 & -\frac{1}{4} & -\frac{1}{4} \\
 \frac{1}{256 \kappa ^2+1} & -\frac{1}{4} & 0 & -\frac{1}{4} \\
 \frac{1}{256 \kappa ^2+1} & -\frac{1}{4} & -\frac{1}{4} & 0 \\
\end{array}
\right)
\end{equation}
And the relevant channel is the same as the case of the first quadrant, but with a smaller eigenvalue,
\begin{align}
    u_{\al,i}=g_{\al}(3,1,1,1)^T, \quad \frac{d g_{\al}}{d \ell} = 1.70 g_{\al}.
\end{align}

\vbox{}
\noindent\textbf{The $r_1$ axis, transition between the $\Oo(4)$ DQCP and first-order transition}: The boson corresponding to the triplet mass is critical, $\{M^a\}=\{\sigma^3\otimes \dsi_N\}$. There are mixture between $V^0,V^3$ and also between $V^1,V^2$, we will present the RG equation for $V^0,V^3$ and $V^1,V^2$ separately. For $\{V^0,V^3\}$,
\begin{align}
    &\frac{d u_{\al,i}}{d \ell} =\left(-1+\frac{64}{3\pi^2 (2N)} \mM_{(\al,i),(\be,j)}\right)u_{\be,j},\\
    &\mM=\left(
\begin{array}{cccc}
 \left(
\begin{array}{cc}
 \frac{4 \left(512 \kappa ^2-1\right)}{\left(256 \kappa ^2+1\right)^2}-\frac{1}{2} & 0 \\
 0 & \frac{2}{256 \kappa ^2+1}-\frac{1}{2} \\
\end{array}
\right) & (256 \kappa ^2+1)^{-1}\dsi_2 & (256 \kappa ^2+1)^{-1}\dsi_2 & (256 \kappa ^2+1)^{-1}\dsi_2 \\
 (256 \kappa ^2+1)^{-1}\dsi_2 & 0_2 & -\frac{1}{4}\sigma^1 & -\frac{1}{4}\sigma^1 \\
 (256 \kappa ^2+1)^{-1}\dsi_2 & -\frac{1}{4}\sigma^1 & 0_2 & -\frac{1}{4}\sigma^1 \\
 (256 \kappa ^2+1)^{-1}\dsi_2 & -\frac{1}{4}\sigma^1 & -\frac{1}{4}\sigma^1 & 0_2 \\
\end{array}
\right)
\end{align}
where the $2\times 2$ matrices act on the $\{V^0,V^3\}$ space, $0_2$ is $2\times 2$ matrix with all entries being 0. The only relevant channel is,
\begin{equation}
    u_{\al,i}=g_{(0,3)}((-0.03,0.82),(-0.071,0.32),(-0.071,0.32),(-0.071,0.32))^T, \quad \frac{d g_{(0,3)}}{d \ell} = 1.89 g_{(0,3)}.
\end{equation}
For $\{V^1,V^2\}$,
\begin{align}
    &\frac{d u_{\al,i}}{d \ell} =\left(-1+\frac{64}{3\pi^2 (2N)} \mM_{(\al,i),(\be,j)}\right)u_{\be,j},\\
    &\mM=\left(
\begin{array}{cccc}
 \left(\frac{2}{256 \kappa ^2+1}+\frac{1}{4}\right) \dsi_2 & (256 \kappa ^2+1)^{-1}\dsi_2-\frac{1}{4}\sigma^1 & (256 \kappa ^2+1)^{-1}\dsi_2-\frac{1}{4}\sigma^1 & (256 \kappa ^2+1)^{-1}\dsi_2-\frac{1}{4}\sigma^1 \\
 (256 \kappa ^2+1)^{-1}\dsi_2-\frac{1}{4}\sigma^1 & -\frac{1}{4}\dsi_2 & 0_2 & 0_2 \\
 (256 \kappa ^2+1)^{-1}\dsi_2-\frac{1}{4}\sigma^1 & 0_2 & -\frac{1}{4}\dsi_2 & 0_2 \\
 (256 \kappa ^2+1)^{-1}\dsi_2-\frac{1}{4}\sigma^1 & 0_2 & 0_2 & -\frac{1}{4}\dsi_2 \\
\end{array}
\right).
\end{align}
There are two relevant channels,
\begin{align}
    u_{\al,i}&=g^{(1)}_{(1,2)}((-3, 3), (-1, 1), (-1, 1), (-1, 1))^T, \quad \frac{d g^{(1)}_{(1,2)}}{d \ell} = 2.78 g^{(1)}_{(1,2)}, \\
    u_{\al,i}&=g^{(2)}_{(1,2)}((4.1,4.1),(1,1),(1,1),(1,1))^T, \quad \frac{d g^{(2)}_{(1,2)}}{d \ell} = 2.02 g^{(2)}_{(1,2)}.
\end{align}
The first channel is antisymmetric combination of $V^1,V^2$ and the second is symmetric combination.

\vbox{}
\noindent\textbf{The origin, multi-critical point}: Both bosons are critical, the boson-fermion vertices are $\{M^a\}=\{\dsi_{2N},\sigma^3\otimes \dsi_N\}$. Again, there will be mixture between $V^0,V^3$ and also between $V^1,V^2$. For $\{V^0,V^3\}$,
\begin{align}
    &\frac{d u_{\al,i}}{d \ell} =\left(-1+\frac{64}{3\pi^2 (2N)} \mM_{(\al,i),(\be,j)}\right)u_{\be,j},\\
    &\mM=\left(
\begin{array}{cccc}
 \left(
\begin{array}{cc}
 \frac{4 \left(512 \kappa ^2-1\right)}{\left(256 \kappa ^2+1\right)^2}-1 & 0 \\
 0 & \frac{2}{256 \kappa ^2+1}-1 \\
\end{array}
\right) & (256 \kappa ^2+1)^{-1}\dsi_2 & (256 \kappa ^2+1)^{-1}\dsi_2 & (256 \kappa ^2+1)^{-1}\dsi_2 \\
 (256 \kappa ^2+1)^{-1}\dsi_2 & 0_2 & -\frac{1}{4}(\dsi_2+\sigma^1) & -\frac{1}{4}(\dsi_2+\sigma^1) \\
 (256 \kappa ^2+1)^{-1}\dsi_2 & -\frac{1}{4}(\dsi_2+\sigma^1) & 0_2 & -\frac{1}{4}(\dsi_2+\sigma^1) \\
 (256 \kappa ^2+1)^{-1}\dsi_2 & -\frac{1}{4}(\dsi_2+\sigma^1) & -\frac{1}{4}(\dsi_2+\sigma^1) & 0_2 \\
\end{array}
\right).
\end{align}
And the relevant channel is the same as previous case with a smaller eigenvalue,
\begin{equation}
    u_{\al,i}=g_{(0,3)}((-0.03,0.82),(-0.071,0.32),(-0.071,0.32),(-0.071,0.32))^T, \quad \frac{d g_{(0,3)}}{d \ell} = 1.35 g_{(0,3)}.
\end{equation}
For $\{V^1,V^2\}$,
\begin{align}
    &\frac{d u_{\al,i}}{d \ell} =\left(-1+\frac{64}{3\pi^2 (2N)} \mM_{(\al,i),(\be,j)}\right)u_{\be,j},\\
    &\mM=\left(
\begin{array}{cccc}
 \left(\frac{2}{256 \kappa ^2+1}-\frac{1}{4}\right) \dsi_2 & (256 \kappa ^2+1)^{-1}\dsi_2-\frac{1}{4}\sigma^1 & (256 \kappa ^2+1)^{-1}\dsi_2-\frac{1}{4}\sigma^1 & (256 \kappa ^2+1)^{-1}\dsi_2-\frac{1}{4}\sigma^1 \\
 (256 \kappa ^2+1)^{-1}\dsi_2-\frac{1}{4}\sigma^1 & -\frac{1}{4}\dsi_2 & -\frac{1}{4}\dsi_2 & -\frac{1}{4}\dsi_2 \\
 (256 \kappa ^2+1)^{-1}\dsi_2-\frac{1}{4}\sigma^1 & -\frac{1}{4}\dsi_2 & -\frac{1}{4}\dsi_2 & -\frac{1}{4}\dsi_2 \\
 (256 \kappa ^2+1)^{-1}\dsi_2-\frac{1}{4}\sigma^1 & -\frac{1}{4}\dsi_2 & -\frac{1}{4}\dsi_2 & -\frac{1}{4}\dsi_2 \\
\end{array}
\right).
\end{align}
There are two relevant channels,
\begin{align}
    u_{\al,i}&=g^{(1)}_{(1,2)}((-3, 3), (-1, 1), (-1, 1), (-1, 1))^T, \quad \frac{d g^{(1)}_{(1,2)}}{d \ell} = 2.24 g^{(1)}_{(1,2)}, \\
    u_{\al,i}&=g^{(2)}_{(1,2)}((4.1,4.1),(1,1),(1,1),(1,1))^T, \quad \frac{d g^{(2)}_{(1,2)}}{d \ell} = 1.49 g^{(2)}_{(1,2)}.
\end{align}
The first relevant channel is the antisymmetric combination of $V^1,V^2$, it is interesting that this relevant channel has the same scaling dimension as the relevant channel $V^3$ with $(3,1,1,1)^T$ in the $\Oo(4)$ DQCP (Eq.~\ref{eq:O(4)_scaling_v3}).


\subsection{Mass scaling}
Combining the diagrams in previous sections allows us to calculate the scaling dimension for the fermion mass term, which corresponds to the vertex $\bpsi V^\al \otimes \Ga^{(0)} \psi\equiv \bpsi V^\al \otimes \mathds{1}_2 \psi$. As discussed in the main text, we use the $N_f=2$ QED$_3$ description of DQCP, and consider its large-$N$ generalization. The vertex of singlet mass is thus $V^\al=\dsi_{2N}$ and for the triplet mass is $V^\al=\sigma^3\otimes \dsi_N$. The diagram equation for the corrections of the mass scaling dimension is,
\begin{fmffile}{mass-scaling}
\begin{equation}
\begin{split}
    \begin{gathered}
        \begin{fmfgraph*}(90,40)
        \fmfleft{i} \fmfright{o} \fmf{fermion}{i,v1,v2,v3,o} \fmf{dbl_wiggly,right,tension=0.3}{v2,v3}
        \phov{v2,v3} \intv{v1}
        \end{fmfgraph*}
    \end{gathered} +\quad
    \begin{gathered}
        \begin{fmfgraph*}(90,40)
        \fmfleft{i} \fmfright{o} \fmf{fermion}{i,v1,v2,v3,o} \fmf{dbl_wiggly,right,tension=0.3}{v1,v3}
        \phov{v1,v3} \intv{v2}
        \end{fmfgraph*}
    \end{gathered} +\quad
    \begin{gathered}
        \begin{fmfgraph*}(90,50)
        \fmfstraight
        \fmfleft{i0,i1,i2} \fmfright{o0,o1,o2}
        \fmf{fermion}{i0,v0,w0,o0}
        \fmf{phantom}{i1,v1,w1,o1}
        \fmffreeze
        \fmfpoly{shaded}{v1,w1,t1}
        \fmf{dbl_wiggly}{v0,v1}
        \fmf{dbl_wiggly}{w0,w1}
        \intv{t1}  \phov{v0,v1,w0,w1}
        \end{fmfgraph*}
    \end{gathered}\\
    +\quad \begin{gathered}
        \begin{fmfgraph*}(90,40)
        \fmfleft{i} \fmfright{o} \fmf{fermion}{i,v1,v2,v3,o} \fmf{dbl_dashes,right,tension=0.3}{v2,v3}
        \phov{v2,v3} \intv{v1}
        \end{fmfgraph*}
    \end{gathered} +\quad
    \begin{gathered}
        \begin{fmfgraph*}(90,40)
        \fmfleft{i} \fmfright{o} \fmf{fermion}{i,v1,v2,v3,o} \fmf{dbl_dashes,right,tension=0.3}{v1,v3}
        \phov{v1,v3} \intv{v2}
        \end{fmfgraph*}
    \end{gathered}
\end{split}
\end{equation}
\end{fmffile}
For $M^a$ being full rank, $\mathbb{M}=2N$, $C_M$ equals to the number of critical boson $N_b$. For singlet mass term, $C_{M,\al} = N_b$, but for the triplet mass, $C_{M,\al}$ depends on the choices of $M^a$, the coefficient is calculated explicitly by $C_{M,\al}=\tr[(\sum_a M^a V^\al M^a)V^\al]/\tr[V^\al V^\al]$. For boson associated to singlet mass, the result is simple, $M^a=\mathds{1}$, with $a=1,...,N_b$, $C_{M,\al}=N_b$.

For the mass scaling, $m=0$, $C_{\ga,(0),\{\}}=3$, and there is no two-loop correction by critical boson. Collecting the logarithmic divergent part, we get,
\begin{align}
    \Delta_{\bar{\psi}\mathds{1}_{2N}\psi} &=2- \frac{128(512\ka^2-1)}{3\pi^2(2N)(1+256\ka^2)^2}+\frac{16 N_b}{3\pi^2 (2N)}\\
    \Delta_{\bar{\psi}(\sigma^3\otimes \dsi_N)\psi} &= 2-\frac{64}{3\pi^2(2N)(1+256\ka^2)}+\frac{4(3C_{M,\al}+N_b)}{3\pi^2 (2N)}
\end{align}
where the last term in each equation comes from the critical boson contribution. This general result agrees with previous work with certain parameters.

\subsection{Boson mass scaling}
We can also calculate the scaling dimension of the boson operator $\phi_a^2$. Following \refcite{boyack_deconfined_2019}, we define the scalar two-point function as $G^{\phi}_{ab}\equiv \langle \phi_a(p)\phi_b(-p) \rangle$, and its $\mathcal{O}(1/N)$ 1PI scalar self-energy contribution is represented by $\Sigma_{ab}^{\phi(1)}(p)$. From the Dyson’s equation, the two-point function to $\mathcal{O}(1/N)$ is,
\begin{equation}
    G^{\phi}_{ab}=D_{ab}(p)+D_{ac}(p)\Sigma_{cd}^{\phi(1)}(p)G^\phi_{db}(p) \simeq D_{ab}(p)+D_{ac}(p)\Sigma_{cd}^{\phi(1)}(p)D_{db}(p)
\end{equation}
where the self-energy is obtained by summing over the basic diagrams for fermion mass scaling but with nontrivial choices of $M^a$s. Because of the coupling $\phi_a \bpsi M^a \psi$, the self-energy corrections depend on $M^a$s and can therefore change the scaling dimensions of the corresponding bosons. The self-energy has the following generic form,
\begin{equation}
    \Sigma_{ab}^{\phi(1)}(p) = \delta_{ab} \frac{c_a\abs{p}}{\pi^2 N}\ln{\left(\frac{\Lambda^2}{p^2}\right)}
\end{equation}
For example, for $M^a = \mathds{1}$, $c= \frac{2}{3} -\frac{16 \left(512 \kappa ^2-1\right)}{3 \left(256 \kappa ^2+1\right)^2}\kaequal 6$, and for $M^a$ is traceless, $c =\frac{2}{3}-\frac{8}{3 \left(256 \kappa ^2+1\right)} \kaequal -2$. The self-energy will contribute to the scaling dimension of the $\phi^2$ in the following diagram,
\begin{fmffile}{boson-self-energy-cor}
\begin{equation}\label{eq:boson_selfE}
    \begin{gathered}
        \begin{fmfgraph*}(60,20)
        \fmfbottom{b1,b2}  \fmftop{v1}
        \fmfstraight
        \fmf{dbl_dashes}{b1,v1,v2,b2}
        \fmfblob{.18w}{v2}
        \bsv{v1}
        \end{fmfgraph*}
    \end{gathered}
\end{equation}
\end{fmffile}
\noindent where the shaded bubble is the self-energy correction $\Sigma_{ab}^{\phi(1)}(p)$. There is one more diagram at $\mathcal{O}(1/N)$ will contribute to the scaling of $\phi^2$, as following,
\begin{fmffile}{boson-box-cor}
\begin{equation}\label{eq:boson_box}
    \begin{gathered}
        \begin{fmfgraph*}(60,60)
        \fmfbottom{b0,b1,b2,b3}  \fmftop{t1}
        \fmfpolyn{tension=1/3}{z}{4}
        \fmf{dbl_dashes}{b0,z4}
        \fmf{dbl_dashes}{b3,z1}
        \fmf{dbl_dashes}{z2,t1,z3}
        \bsv{t1}
        \end{fmfgraph*},
    \end{gathered}
\end{equation}
\end{fmffile}
\noindent the fermion ``box'' is the summation of fermions running clockwise and anti-clockwise. The scaling dimension of $\phi^2$ is combining Eq.~\ref{eq:boson_selfE} and Eq.~\ref{eq:boson_box}, this gives,
\begin{equation}
    \Delta_{\phi_a^2}=2-\frac{16c_a}{\pi^2 N}+\frac{8}{\pi^2 N}.
\end{equation}
Note that the hourglass diagram (the first diagram in Eq.~\ref{eq:boson_hourglass}) won't contribute to the anomalous dimension, a simple argument is that similar diagram with one internal boson line appears in the self-energy correction (second and third diagram in Eq.~\ref{eq:boson_hourglass}) and it contributes to the anomalous dimension, while the hourglass diagram has two internal boson line, the power in the denominator is larger by 1, hence, it won't contribute to the anomalous dimension.
\begin{fmffile}{boson-hourglass}
\begin{equation}\label{eq:boson_hourglass}
    \begin{gathered}
        \begin{fmfgraph*}(60,40)
        \fmfleft{i} \fmfright{o}
        \fmfpolyn{tension=0}{z}{4}
        \fmf{dbl_dashes}{i,z3}
        \fmf{dbl_dashes}{o,z1}
        \fmf{dbl_dashes}{z2,v,z4}
        \bsv{v}
        \end{fmfgraph*}
    \end{gathered}, \quad
    \begin{gathered}
        \begin{fmfgraph*}(60,40)
        \fmfleft{i} \fmfright{o}
        \fmfpolyn{tension=0}{z}{4}
        \fmf{dbl_dashes}{i,z3}
        \fmf{dbl_dashes}{o,z1}
        \fmf{dbl_dashes}{z2,z4}
        \end{fmfgraph*}
    \end{gathered}
    \  \subset\
    \begin{gathered}
        \begin{fmfgraph*}(60,40)
        \fmfleft{i} \fmfright{o}
        \fmfstraight
        \fmf{dbl_dashes}{i,v,o}
        \fmfblob{.3w}{v}
        \end{fmfgraph*}
    \end{gathered}
\end{equation}
\end{fmffile}
With $\mathcal{O}(1/N)$ correction, the scaling dimension of the boson operator $\phi_a^2$ are listed in the Tabel.~\ref{tab:scaling_boson}. These scaling dimensions are not trustworthy for small fermion flavors $N$, but they show a trend for the scaling dimensions when having different boson-fermion vertices in large $N$.
\begin{table}[]
    \centering
    \begin{tabular}{c|c}
        $\{M^a\}$ & $\Delta_{\phi_a^2}$ \\ \hline
        $\{\dsi_2\}$ & $2-\frac{8}{3 \pi ^2 N}+\frac{256 \left(512 \kappa ^2-1\right)}{3 \pi ^2 \left(256 \kappa ^2+1\right)^2 N} \kaequal 2-\frac{88}{\pi ^2 N}$ \\\hline
        $\{\sigma^3\}$ & $2-\frac{8}{3 \pi ^2 N}+\frac{128}{3 \pi ^2 \left(256 \kappa ^2+1\right) N}\kaequal 2+\frac{40}{\pi ^2 N}$\\\hline
        $\{\dsi_2,\sigma^3\}$ & $\{2-\frac{40}{3 \pi ^2 N}+\frac{256 \left(512 \kappa ^2-1\right)}{3 \pi ^2 \left(256 \kappa ^2+1\right)^2 N},2-\frac{40}{3 \pi ^2 N}+\frac{128}{3 \pi ^2 \left(256 \kappa ^2+1\right) N}\} \kaequal \{2-\frac{296}{3 \pi ^2 N}, 2+\frac{88}{3 \pi ^2 N}\}$
    \end{tabular}
    \caption{The scaling dimensions of $\phi^2_a$ with several choices of the boson-fermion vertices, these choices correspond to the axis and origin of the phase diagram Fig.~\ref{fig:phase_diagram_multi} in the main text.}
    \label{tab:scaling_boson}
\end{table}

\section{Details of \ntqed and self-duality}\label{sec:app_self-dual}
The single flavor fermion coupled to the $\U(1)$ gauge field is dual to free fermion theory and this is dubbed as fermion/fermion duality\cite{hsin_levelrank_2016,seiberg_a-duality_2016},
\begin{align}
    i \bar{\Psi} \slashed{D}_{A_1} \Psi \Longleftrightarrow i \bar{\chi}\slashed{D}_{a_1} \chi -\frac{2}{4\pi}b_1db_1+\frac{1}{2\pi}a_1db_1+\frac{1}{2\pi}A_1db_1-\frac{1}{4\pi}A_1dA_1 -2CS_g, \label{eq:ff_dual}\\
    i \bar{\Psi} \slashed{D}_{A_2} \Psi \Longleftrightarrow i \bar{\chi}\slashed{D}_{a_2} \chi +\frac{1}{4\pi}a_2da_2+\frac{2}{4\pi}b_2db_2 -\frac{1}{2\pi}a_2db_2-\frac{1}{2\pi}A_2db_2+2CS_g,\label{eq:ff_dual_tr}
\end{align}
where $CS_g$ denotes the gravitational Chern-Simons term which will vanish in the flat spacetimes. The second line is the orientation reversed (time-reversal) version of the first one. We can then product them together on each side with the substitution $A_1 \rightarrow A, A_2 \rightarrow A-2X$. Next, adding the counterterms $\tpi Ad(Y-X)+\fpi(XdX-YdY)+\fpi AdA+2CS_g$ to both sides and gauging $A$, after integrating out most of the gauge fields, we get,
\begin{align}
    &\ii \bar{\Psi}_1 \slashed{D}_{a} \Psi_1+\ii \bar{\Psi}_2 \slashed{D}_{a-2X} \Psi_2 +\fpi ada+\tpi ad(Y-X)+\fpi(XdX-YdY)+2CS_g \Longleftrightarrow\\
     &\ii \bar{\chi}_1 \slashed{D}_{\ta-2Y} \chi_1+\ii \bar{\chi}_2 \slashed{D}_{\ta} \chi_2 +\fpi\ta d\ta+\tpi \ta d(X-Y)+\fpi(YdY-XdX)+2CS_g,
\end{align}
the self-duality exchanges $X\leftrightarrow Y$ and $\chi_i \leftrightarrow \Psi_{\bar{i}}$. After relabeling the dynamical gauge fields $a,\ta$, it gives back \eqnref{eq:selfdual_1} and \eqnref{eq:selfdual_2}.

The self-duality exchanges the monopole symmetry and the Cartan subgroup of the flavor symmetry. It is also the duality between strong and weak couplings, this can be seen from the duality transformations amongst the derivation and their corresponding transformations in the 3+1d bulk. Considering the 2+1d $\U(1)$ gauge matter theories live at the boundary of 3+1d $\U(1)$ gauge theory with the coupling constant $\tau$,
\begin{align}
    I(A)&=\frac{1}{8\pi} \int_X d^4x \sqrt{g}\left(\frac{2\pi}{e^2}F_{mn}F^{mn}+\frac{\ii \theta}{4\pi}\epsilon_{mnpq}F^{mn}F^{pq} \right)\\
    &=\frac{\ii}{8\pi}\int_X d^4x\sqrt{g}(\Bar{\tau}F_{mn}^+F^{+mn}-\tau F_{mn}^-F^{-mn}), \quad \tau = \frac{\theta}{2\pi}+\frac{2\pi\ii}{e^2}
\end{align}
where $g$ is the metric for the spacetime, the theory and the transformation properties are well-defined also in the curved spacetime. $F=dA$ and $F$ is decomposed into self-dual and anti-self-dual pieces, $F_{mn}^\pm = \frac{1}{2}(F_{mn}\pm (\star F)_{mn})$ with $(\star F)_{mn} = \frac{1}{2}\epsilon_{mnpq}F^{pq}$, also $(\star F)_{mn}(\star F)^{mn} = F_{mn}F^{mn}$. The $S$ transformation and $T$ transformation act as,
\begin{align}
    S&:\quad \tau \rightarrow \tau' = -\frac{1}{\tau},\quad \begin{pmatrix} 0 & -1\\1 & 0 \end{pmatrix},\quad \int_{\partial X} J\cdot A\rightarrow \int_{\partial X} J \cdot a-\frac{1}{2\pi} adA'\\
    -S&:\quad \tau \rightarrow \tau' = -\frac{1}{\tau},\quad \begin{pmatrix} 0 & -1\\1 & 0 \end{pmatrix},\quad \int_{\partial X} J\cdot A\rightarrow \int_{\partial X} J \cdot a+\frac{1}{2\pi} adA'
\end{align}
\begin{equation}
    T[k]:\quad \tau \rightarrow \tau' = \tau+k, \quad \begin{pmatrix} 1 & k \\ 0 & 1 \end{pmatrix},\quad \int_{\partial X} J\cdot A \rightarrow \int_{\partial X} J\cdot A-\frac{k}{4\pi}AdA
\end{equation}
The $SL(2,\Z)$ matrix acts on the coupling constant $\tau$ as,
\begin{equation}
    \tau \rightarrow \tau'=\frac{a \tau+b}{c\tau+d}, \quad \begin{pmatrix} a & b\\c & d \end{pmatrix} \in SL(2,\Z)
\end{equation}
The fermion/fermion duality in the derivation of the \ntqed self-duality is essential in connecting the left-hand-side and the right-hand-side (the other procedures, adding the counterterms and gauging the background gauge fields are the same for both hand sides). Using the above notation, the fermion/fermion duality and its orientation reversed version is,
\begin{align}
    &T[1]\circ (-S)\circ T[2]\circ(-S),\ \tau \rightarrow \frac{1}{2}-\frac{1}{2(2\tau-1)}\\
    &S\circ T[-2]\circ S\circ T[-1],\ \tau \rightarrow \frac{1}{2}-\frac{1}{2(2\tau-1)}.
\end{align}
Take the coupling of the bulk theory $\tau =\frac{1}{2}+\frac{2\pi\ii}{e^2}$, under the duality $\tau\rightarrow \frac{1}{2}-\frac{1}{2(2\tau-1)} = \frac{1}{2}+\frac{e^2\ii}{8\pi}$. If $e\rightarrow 0$, which is the weak coupling limit, the dual theory has the strong coupling with $\tau\rightarrow\frac{1}{2}+0\ii$. This suggests the fermion/fermion duality is a strong-weak duality, and similar calculation can be done for the \ntqed, which involves the $\U(1)\times \U(1)$ gauge theory in the bulk.

\section{Connection to the gapless $\Z_2$ spin liquid in \refcite{shackleton2021deconfined}}\label{sec:app_z2_spin_liquid}
\subsection{Matrix form of fermion operators}
The 2 flavor Nambu spinor can be written in matrix form,
\begin{equation}
    \CX_i=\begin{pmatrix}f_{i\uparrow} & -f_{i\downarrow}^\dagger \\
    f_{i\downarrow} & f_{i\uparrow}^\dagger\end{pmatrix}
\end{equation}
The $\gSU(2)$ gauge symmetry and physical spin symmetry act as,
\begin{align}
    &\gSU(2)_g: \CX_i\rightarrow \CX_i U_{g,i}^\dagger \\
    &\gSU(2)_s: \CX_i\rightarrow U_{s,i}\CX_i.
\end{align}
In majorana basis, one has,
\begin{equation}
    \CX_i=\frac{1}{\sqrt{2}}(\chi_0+\ii \chi_a \sigma^a).
\end{equation}
Note that there is a discrepancy in the conventional notation and this, but it is merely relabeling,
\begin{equation}
    \begin{pmatrix}
        f_\uparrow \\ f_\uparrow^\dagger \\ f_\downarrow \\ f_\downarrow^\dagger
    \end{pmatrix} = \begin{pmatrix}
    1 & \ii & 0 & 0 \\
    1 & -\ii & 0 & 0 \\
    0 & 0 & 1 & \ii \\
    0 & 0 & 1 & -\ii
    \end{pmatrix}\begin{pmatrix}
        \chi_{1,1}\\ \chi_{1,2}\\ \chi_{2,1}\\ \chi_{2,2}\\
    \end{pmatrix}
    =\begin{pmatrix}
    1 & 0 & 0 & \ii \\
    1 & 0 & 0 & -\ii \\
    0 & \ii & -1 & 0 \\
    0 & -\ii & -1 & 0
    \end{pmatrix}\begin{pmatrix}
        \chi_{0}\\ \chi_{1}\\ \chi_{2}\\ \chi_{3}\\
    \end{pmatrix}
\end{equation}
the relabeling is,
\begin{equation}\label{eq:relabel}
    \begin{pmatrix}
        \chi_{0}\\ \chi_{1}\\ \chi_{2}\\ \chi_{3}\\
    \end{pmatrix} = \begin{pmatrix}
     1 & 0 & 0 & 0 \\
 0 & 0 & 0 & 1 \\
 0 & 0 & 1 & 0 \\
 0 & 1 & 0 & 0
    \end{pmatrix}\begin{pmatrix}
        \chi_{1,1}\\ \chi_{1,2}\\ \chi_{2,1}\\ \chi_{2,2}\\
    \end{pmatrix}.
\end{equation}
$\chi_a\rightarrow \chi_{a_1,a_2}$.
\subsection{Hamiltonian and Higgs fields}
Define the $4 \times 2$ matrix operator,
\begin{align}
    &X_{\alpha,v;\beta}=\frac{1}{\sqrt{2}}(\chi_{0,v}\dsi_{\alpha \beta }+\ii\chi_{a,v}\sigma^a_{\alpha \beta}) \\
    &X_v=\frac{1}{\sqrt{2}}(\chi_{0,v}\sigma^0+\ii\chi_{a,v}\sigma^a)
\end{align}
with $\chi_{a,v}, a=0\sim 3, v=1,2$. The $\gamma$-matrices act on the spinor index $m$ in $\chi_{m,a,v}$, and it is left implicit. The mean-field Lagrangian is,
\begin{align*}
    \CL&=\ii \Tr{\bar{X}\gamma^\mu\partial_\mu X}\\
    &=\ii \Tr{(\chi^T_{0,v}\gamma^0\sigma^0-\ii\chi^T_{a,v}\gamma^0\sigma^a)\gamma^\mu\partial_\mu (\chi_{0,v}\sigma^0+\ii\chi_{b,v}\sigma^b)}\\
    &=\sum_{a,v} \ii \chi^T_{a,v}\gamma^0\gamma^\mu\partial_\mu\chi_{a,v}
\end{align*}
where $\gamma^\mu=\{\sigma^2,\sigma^3,\sigma^1\}$ and $\bar X=X^\dagger \gamma^0$.

Let's now proceed to translate the Lagrangian for the Higgs fields in \refcite{shackleton2021deconfined}, the matrix $\mu^i$ acts on the $v$ indices, one of the $\IZ_2$ Higgs field is,
\begin{align}
    &\Phi_1^a\Tr{\sigma^a\bar{X}\mu^z\gamma^x X}\nonumber\\
    =&\Phi_1^c\Tr{\sigma^c(\chi^T_{0,v}\gamma^0\sigma^0-\ii\chi^T_{a,v}\gamma^0\sigma^a)\mu^z_{v,w}\gamma^x (\chi_{0,w}\sigma^0+\ii\chi_{b,w}\sigma^b)}\nonumber\\
    =&\Phi_1^c\ii(\chi^T_{0,v}\gamma^{0}\gamma^x\mu^z_{v,w}\chi_{c,w}-\chi^T_{c,v}\gamma^0\gamma^x\mu^z_{v,w}\chi_{0,w})
\end{align}
one can also get the matrices that act on the index $a$,
\begin{align}
    &c=1, \delta_{i,0}\delta_{j,1}-\delta_{i,1}\delta_{j,0} = -\ii(\sigma^{02}+\sigma^{32})\\
    &c=2, \delta_{i,0}\delta_{j,1}-\delta_{i,1}\delta_{j,0} = -\ii(\sigma^{20}+\sigma^{23})\\
    &c=3, \delta_{i,0}\delta_{j,1}-\delta_{i,1}\delta_{j,0} = -\ii(\sigma^{12}+\sigma^{21})
\end{align}
To compare with our model, we need to change the basis following Eq.~\ref{eq:relabel},
\begin{align}
    &c=1, -\ii M^1=-\ii(\sigma^{12}+\sigma^{21})\\
    &c=2, -\ii M^2=-\ii(\sigma^{20}+\sigma^{23})\\
    &c=3, -\ii M^3=-\ii(\sigma^{02}+\sigma^{32})
\end{align}
and using the basis $\chi_{m,v,a_1,a_2}$, therefore, the Higgs field becomes,
\begin{equation}
    \Phi_1^c \chi^T[(\gamma^{0}\gamma^x)\otimes \mu^z\otimes M^c ]\chi
\end{equation}
the other $\IZ_2$ Higgs field is,
\begin{equation}
    \Phi_2^c \chi^T[(\gamma^{0}\gamma^y)\otimes \mu^x\otimes M^c ]\chi
\end{equation}
and the $\gU(1)$ Higgs field is,
\begin{equation}
    \Phi_3^c \chi^T[\gamma^{0}(\gamma^y k_x +\gamma^x k_y)\otimes \mu^y\otimes M^c ]\chi
\end{equation}
\subsection{The Higgs configuration}
\refcite{shackleton2021deconfined} proposes the staggered flux state is obtained when $\langle \Phi_3 \rangle\propto (0,0,\delta \phi)$ and the $Z2Azz13$ state follows from $\langle \Phi_1 \rangle \propto (\gamma_1-\gamma_2,\gamma_1+\gamma_2,0)$ and $\langle \Phi_2 \rangle \propto (-\gamma_1-\gamma_2,\gamma_1-\gamma_2,0)$. Recall that, $\gamma^\mu=\{\sigma^2,\sigma^3,\sigma^1\}$ and
\begin{align}
    &c=1, -\ii M^1=-\ii(\sigma^{12}+\sigma^{21})\\
    &c=2, -\ii M^2=-\ii(\sigma^{20}+\sigma^{23})\\
    &c=3, -\ii M^3=-\ii(\sigma^{02}+\sigma^{32}).
\end{align}
When condensing the Higgs fields, it corresponds to generate the mass for the combination of the fermion bilinears,
\begin{align}\label{eq:higgs_fields}
    \Phi_1^{1,2}:\sigma^{1312}+\sigma^{1321},\sigma^{1320}+\sigma^{1323}\\
    \Phi_2^{1,2}:\sigma^{3112}+\sigma^{3121},\sigma^{3120}+\sigma^{3123}\\
    \Phi_3^3:\sigma^{1202}k_y+\sigma^{1232}k_y,\sigma^{3202}k_x+\sigma^{3232}k_x
\end{align}
and the kinetic terms are,
\begin{equation}
    \sigma^{1000}k_x,\sigma^{3000}k_y.
\end{equation}
The only Pauli matrix commutes with the above matrices is $\sigma^{0230}$, which is also the symmetry generator.

\paragraph{Our model} In our model, the kinetic terms are,
\begin{equation}
    \sigma^{100}k_x, \sigma^{300}k_y
\end{equation}
and the pairing terms are
\begin{equation}
    \sigma^{323},\sigma^{321},\sigma^{123},\sigma^{121}
\end{equation}
and the Pauli matrices that commute with the above are,
\begin{equation}
    \sg{012},\sg{020},\sg{032}
\end{equation}
\subsection{Basis rotation}
We can match both theories by examining their symmetry generators. The only matrix $\sg{0230}$ that commutes with other matrices in \refcite{shackleton2021deconfined} can be rotated to,
\begin{align}
    &\sg{0012}, \text{ by }\cfr{0222}\\
    &\sg{0020}, \text{ by }\cfr{0210}\\
    &\sg{0332}, \text{ by }\cfr{0102}
\end{align}
where the rotation is generated by $\sigma^I\rightarrow e^{-\ii\frac{\pi}{4}\sigma^J}\sigma^{I} e^{\ii\frac{\pi}{4}\sigma^J}$. The $\IZ_2$ Higgs fields in \eqnref{eq:higgs_fields} will be rotated to,
\begin{align}
    &\Phi_1^{1,2}:\sigma^{1312}+\sigma^{1321},\sigma^{1102}+\sigma^{1323}\nonumber\\
    &\Phi_2^{1,2}:\sigma^{3112}+\sigma^{3121},-\sigma^{3302}+\sigma^{3123} \text{ by } \cfr{0222}
\end{align}
\begin{align}
    &\Phi_1^{1,2}:\sigma^{1102}+\sigma^{1321},\sigma^{1320}+\sigma^{1323}\nonumber\\
    &\Phi_2^{1,2}:-\sigma^{3302}+\sigma^{3121},\sigma^{3120}+\sigma^{3123} \text{ by } \cfr{0210}
\end{align}
\begin{align}\label{eq:match_rot}
    &\Phi_1^{1,2}:-\sigma^{1210}+\sigma^{1321},-\sigma^{1222}+\sigma^{1323}\nonumber\\
    &\Phi_2^{1,2}:\sigma^{3112}-\sigma^{3023},\sigma^{3120}+\sigma^{3021} \text{ by } \cfr{0102}
\end{align}
If one takes the second index as labeling the original theory and the dual theory in our model, some terms of the $\IZ_2$ Higgs fields in \refcite{shackleton2021deconfined} correspond to the pairing in the form of $\sg{121},\sg{123}, \sg{321},\sg{323}$ that appear in both the original theory and the dual theory according to the \eqnref{eq:match_rot}. For example,
\begin{align}
    ...+\chi^T \sg{1321}\chi =...+ \chi_1^T \sg{121}\chi_1 -\chi_2^T \sg{121}\chi_2\sim...+ \psi^\TT \si^2 \ga^0\ga^x \psi-\tpsi^\TT \si^2 \ga^0\ga^x \tpsi \\
    ...+\chi^T \sg{3021}\chi =...+ \chi_1^T \sg{321}\chi_1 +\chi_2^T \sg{321}\chi_2\sim...+ \psi^\TT \si^2 \ga^0\ga^y \psi+\tpsi^\TT \si^2 \ga^0\ga^y \tpsi
\end{align}
where $\psi$ is the original fermion and $\tpsi$ is the dual fermion, they are corresponding to the pairing fermion bilinears that appear in the $\Z_2$ Higgs fields \eqnref{eq:match_rot}. However, the dual fermion pairings are not explicit in the self-dual \ntqed theory and the linear combinations with another fermion bilinears are crucial to obtain the $\Z_2$ Higgs fields in \refcite{shackleton2021deconfined}, for example, $\Phi_1^1 \chi^T(-\sigma^{1210}+\sigma^{1321})\chi$ in the first line of \eqnref{eq:match_rot}.
\bibliographystyle{apsrev4-1}
\end{document}